\documentclass[
  journal=pasa,
  manuscript=research-paper, 
  year=2020,
  volume=37,
]{cup-journal}

\usepackage{microtype,siunitx,booktabs}

\usepackage{graphicx}
\usepackage{amsmath}	
\usepackage{amssymb}	
\usepackage{multicol}       
\usepackage{bm}		
\usepackage{pdflscape}	
\usepackage{geometry}
\usepackage{array}
\usepackage{hyperref} 
\usepackage[T1]{fontenc}
\usepackage{ae,aecompl}
\usepackage{xcolor} 
\usepackage{booktabs}
\usepackage{adjustbox} 
\usepackage{pifont}
\usepackage{nicematrix}

\title[Radio Galaxy Catalogue for EMU-PS]{RG-CAT: Detection Pipeline and Catalogue of Radio Galaxies in the EMU Pilot Survey}

\author{Nikhel Gupta$^{1}$}
\author{Ray P. Norris$^{2,3}$}
\author{Zeeshan Hayder$^{4}$} 
\author{Minh Huynh$^{1,5}$}
\author{Lars Petersson$^{4}$}
\author{X. Rosalind Wang$^{2}$}
\author{Andrew M. Hopkins$^{6,2}$}
\author{Heinz Andernach$^{7}$}
\author{Yjan Gordon$^{8}$}
\author{Simone Riggi$^{9}$}
\author{Miranda Yew$^{2}$}
\author{Evan J. Crawford$^{2}$}
\author{B\"{a}rbel Koribalski$^{3,2}$}
\author{Miroslav D. Filipovi\'c$^{2}$}
\author{Anna D. Kapi\'nska$^{10}$}
\author{Stanislav Shabala$^{11}$}
\author{Tessa Vernstrom$^{5,1}$}
\author{Joshua R. Marvil$^{10}$}
\email[Nikhel Gupta]{Nikhel.Gupta@csiro.au}

\affiliation{
$^1$ CSIRO Space \& Astronomy, PO Box 1130, Bentley WA 6102, Australia \\
$^2$ Western Sydney University, Locked Bag 1797, Penrith, NSW 2751, Australia \\
$^3$ CSIRO Space \& Astronomy, P.O. Box 76, Epping, NSW 1710, Australia \\
$^4$ CSIRO Data61, Black Mountain ACT 2601, Australia \\
$^5$ International Centre for Radio Astronomy Research, The University of Western Australia, 35 Stirling Highway, Crawley, WA 6009, Australia \\ 
$^6$ School of Mathematical and Physical Sciences, 12 Wally’s Walk, Macquarie University, NSW 2109, Australia \\
$^7$ Th\"uringer Landessternwarte, Sternwarte 5, Tautenburg, D-07778 Tautenburg, Germany;
on leave of absence from: Departamento de Astronom\'{i}a, Universidad de Guanajuato, Callej\'on de Jalisco s/n, Guanajuato 36023, GTO, Mexico \\
$^8$ Department of Physics, University of Wisconsin-Madison, 1150 University Avenue, Madison, WI 53706, USA \\
$^9$ INAF-Osservatorio Astrofisico di Catania, Via Santa Sofia 78, 95123, Catania, Italy \\
$^{10}$ National Radio Astronomy Observatory, PO Box 0, Socorro NM, 87801, USA \\
$^{11}$ School of Natural Sciences, University of Tasmania, Private Bag 37, Hobart 7001, Australia \\
}

\received {dd Mmm YYYY}
\revised  {dd Mmm YYYY}
\accepted {dd Mmm YYYY}
\published{dd Mmm YYYY}


\usepackage{amsmath,amsfonts,amssymb}
\usepackage{color}
\usepackage{mathrsfs}
\usepackage{tabularx}
\usepackage{floatrow}
\usepackage{float}

\definecolor{ored}{rgb}{1.00,0.27,0.00}
\definecolor{mygreen}{rgb}{0.2,0.7,0.2}

\definecolor{Gray}{gray}{0.5}
\definecolor{LightCyan}{rgb}{0.88,1,1}

\def \BE{\begin{equation}}
\def \EE{\end{equation}}	
\def \BC{\begin{center}}
\def \EC{\end{center}}
\def \BEA{\begin{eqnarray}}
\def \EEA{\end{eqnarray}}

\def \SIGMA8{\sigma_{8}}

\usepackage{cellspace}
\begin{document}\sloppy\sloppypar\raggedbottom\frenchspacing


\begin{abstract}
We present source detection and catalogue construction pipelines to build the first catalogue of radio galaxies from the 270 $\rm deg^2$ pilot survey of the Evolutionary Map of the Universe (EMU-PS) conducted with the Australian Square Kilometre Array Pathfinder (ASKAP) telescope.
The detection pipeline uses Gal-DINO computer-vision networks \citep{gupta2023b} to predict the categories of radio morphology and bounding boxes for radio sources, as well as their potential infrared host positions. 
The Gal-DINO network is trained and evaluated on approximately 5,000 visually inspected radio galaxies and their infrared hosts, encompassing both compact and extended radio morphologies.
We find that the Intersection over Union (IoU) for the predicted and ground truth bounding boxes is larger than 0.5 for 99\% of the radio sources, and 98\% of predicted host positions are within $3^{\prime \prime}$ of the ground truth infrared host in the evaluation set.
The catalogue construction pipeline uses the predictions of the trained network on the radio and infrared image cutouts based on the catalogue of radio components identified using the \textit{Selavy} source finder algorithm.
Confidence scores of the predictions are then used to prioritize \textit{Selavy} components with higher scores and incorporate them first into the catalogue. 
This results in identifications for a total of 211,625 radio sources, with 201,211 classified as compact and unresolved.
The remaining 10,414 are categorized as extended radio morphologies, including 582 FR-I, 5,602 FR-II, 1,494 FR-x (uncertain whether FR-I or FR-II), 2,375 R (single-peak resolved) radio galaxies, and 361 with peculiar and other rare morphologies.
Each source in the catalogue includes a confidence score. 
We cross-match the radio sources in the catalogue with the infrared and optical catalogues, finding infrared cross-matches for 73\% and photometric redshifts for 36\% of the radio galaxies.
The EMU-PS catalogue and the detection pipelines presented here will be used towards constructing catalogues for the main EMU survey covering the full southern sky.
\end{abstract}


\section{Introduction}
\label{SEC:Intro}
Radio galaxies remain enigmatic subjects in the realm of astronomy, with much still to be uncovered.
The majority of radio galaxies host Active Galactic Nuclei (AGN), which commonly emit more energy in the radio part of the electromagnetic spectrum than in other wavelengths, such as optical or infrared.
While progress has been made in understanding some aspects, there are key questions that elude us. 
For instance, the precise triggers that activate their powerful radio emission or jets, their interplay with the intergalactic medium, their magnetic field structure, as well as how they influence the broader cosmic environment, remain unresolved.
While in the majority of previous radio surveys, the sources appear as unresolved, with better angular resolution and sensitivity of radio telescopes leads to the detection of a greater number of radio galaxies characterized by intricate, extended structures \citep[see e.g.,][]{norris17a}. 
These structures often consist of multiple components, each displaying distinctive peaks in radio emission. 
The morphologies of equatorially symmetrical extended radio emission from galaxies are broadly classified into two categories: Fanaroff-Riley Class I (FR-I) and Class II (FR-II) type radio galaxies \citep{fanaroff74}. 
These radio galaxies produce highly collimated jets emerging in opposite directions from AGN at the centre of the host galaxy. 
As the distance from the host galaxy increases, the surface brightness of FR-I radio galaxies decreases. 
In contrast, FRII radio galaxies typically feature linear jets terminating in high-brightness hotspots of large radio lobes. 
In some instances, a bipolar jet from an FR-II-type radio galaxy may transition into two distinct radio lobes unconnected by radio emission to the host galaxy.
Consequently, FR-I and FR-II radio galaxies are commonly referred to as edge-darkened and edge-brightened AGN, respectively.

The emergence of new technologies such as phased array feed receivers \citep[PAF;][]{hay06} has enabled swift scanning of extensive portions of the sky, facilitating rapid surveys of large sky areas at radio wavelengths. 
Such advancements have opened up new avenues for detecting millions of radio galaxies. 
For example, the ongoing Evolutionary Map of the Universe \citep[EMU;][]{norris21} survey, conducted with the Australian Square Kilometre Array Pathfinder \citep[ASKAP;][]{johnston07ASKAP,DeBoer09,hotan21} telescope, aims to discover over 40 million compact and extended radio sources within five years \citep{norris21}. 
Similarly, the omnidirectional dipole antennas employed in the Low-Frequency Array \citep[LOFAR;][]{vanharleem13} survey, which spans the entire northern sky, are expected to detect over 15 million radio sources.
Additional cutting-edge radio survey telescopes include Murchison Widefield Array \citep[MWA;][]{wayth18}, MeerKAT \citep{jonas16}, and the Karl G. Jansky Very Large Array \citep[JVLA][]{perley11}.
The Very Large Array Sky Survey \citep[VLASS;][]{lacy20} conducted by JVLA is expected to detect around 5 million radio galaxies.
The forthcoming Square Kilometre Array (SKA\footnote{https://www.skatelescope.org/the-ska-project/}) radio telescope is expected to further escalate the number of galaxy detections, potentially reaching hundreds of millions.
Such a vast dataset will significantly impact our understanding of the physics underlying galaxy evolution. 
In essence, the detection of radio galaxies at various stages of their existence holds immense potential for uncovering concealed dimensions of their behaviour, leading to new insights into their evolutionary processes. 
To fully harness the potential of these radio surveys, it is necessary to redesign the techniques employed in constructing catalogues of radio galaxies.

Grouping the associated components of extended radio galaxies is a necessary step in creating catalogues of radio galaxies. 
This is essential for estimating the actual number density and total integrated flux density of radio galaxies.
Incorrectly grouping associated components or erroneously grouping unassociated components can lead to the misestimation of number density and total flux density, resulting in inaccurate models.
While some analytical approaches are being developed \citep[e.g.,][]{gordon23}, currently, the cross-identification of associated radio galaxy components primarily relies on visual inspections \citep[e.g.,][]{banfield15}.
The widely used source extraction algorithms, such as \textit{Selavy} \citep[][]{whiting12} and AEGEAN \citep[][]{hancock18}, are prone to potential confusion when dealing with detached lobes of extended radio galaxies, as well as neighbouring unassociated compact radio galaxies. 
However, visual inspections are time-consuming, require scientific expertise, and are not scalable to the millions of radio galaxies expected to be discovered in the next few years.

This underscores the vital need to develop computer vision methods for grouping associated radio galaxy components. 
The nature of available data typically determines the trajectory of computer vision tasks, categorizing them into four primary methods: self-supervised, semi-supervised, weakly-supervised and supervised.
Self-supervised learning involves the utilization of unsupervised techniques to train models on the underlying data structure, thereby eliminating the necessity for explicit annotations. 
This has proven effective in discovering new radio morphologies in radio surveys \citep[e.g.][]{galvin20, mostert21, gupta22}. 
Semi-supervised learning combines labelled and unlabeled data during the training process, as demonstrated in the classification of radio galaxies \citep{slijepcevic22}. 
In weakly supervised learning, indirect labels are leveraged for the entire training dataset, serving as a supervisory signal. 
This specific approach has found utility in the classification and detection of extended radio galaxies \citep[][]{gupta2023a}. 
In supervised learning, the model undergoes training using image-label pairs, where these labels provide complete information required for the model to make specific predictions. 
Recently, machine learning (ML) techniques, as exemplified by studies such as e.g \citet[][]{lukic18, alger18, wu19, bowles20, viera21, becker21, brand23, riggi23, sortino23, lao23, gupta2023b}, have found application in the morphological classification and detection of radio galaxies.

This paper builds upon the RadioGalaxyNET dataset \citep[][]{gupta2023c} and computer vision algorithms \citep[][]{gupta2023d} designed to address the challenge of associating radio galaxy components \citep[][]{gupta2023b}. 
The RadioGalaxyNET dataset was curated by professional astronomers through multiple visual inspections and includes multimodal images of the radio and infrared sky, along with pixel-level labels on associated components of extended radio galaxies and their infrared hosts. 
In addition to the 2,800 extended radio galaxies in RadioGalaxyNET, we visually inspected and labelled approximately 2,100 compact radio galaxies and 99 sources with peculiar and other rare morphologies in the present work. 
The annotations comprise class information on the radio-morphological class for these radio galaxies, bounding box information to capture associated components of each radio galaxy, segmentation masks for radio emission, and positions of their host galaxies in infrared images. 
Using this comprehensive dataset of extended and compact radio galaxies and their infrared hosts, we train the Gal-DINO multimodal model \citep[][]{gupta2023b} to simultaneously predict bounding boxes for radio galaxies and potential keypoint positions of their infrared hosts, where a keypoint in machine learning refers to a specific point or landmark in images.
These detections are subsequently employed to generate the first catalogue of compact and extended radio galaxies observed in the first EMU pilot survey (EMU-PS) conducted with the ASKAP telescope.

The structure of the paper is outlined as follows. 
In Section~\ref{SEC:dataset}, we explain the radio and infrared observations, image characteristics, and the labels utilized for training and assessing the computer vision networks. 
Section~\ref{SEC:MLMethods} presents the Gal-DINO network, encompassing training and evaluation specifics, along with the outcomes of network evaluation. 
Details about the catalogue construction pipeline are presented in Section~\ref{SEC:Catalog}. 
Section~\ref{SEC:CatDescription} provides detailed information regarding the catalogue. 
We summarise our findings in Section~\ref{SEC:Summary} and provide directions for future work.

\begin{figure*}
\centering
\vspace*{-2.7cm}
\includegraphics[trim=2.9cm 3cm 2.9cm 1cm, width=8.8cm, scale=0.5]{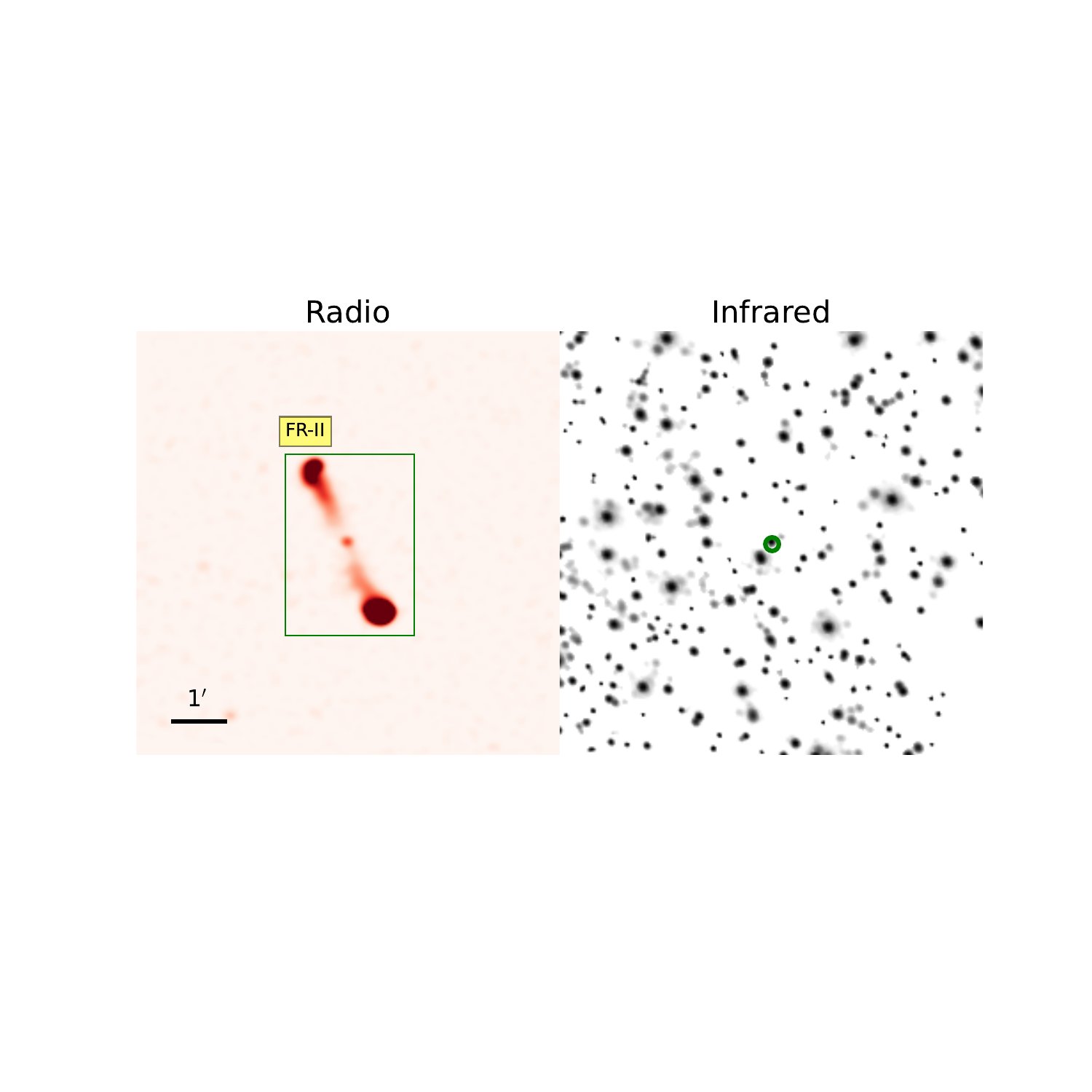}
\includegraphics[trim=2.9cm 3cm 2.9cm 1cm, width=8.8cm, scale=0.5]{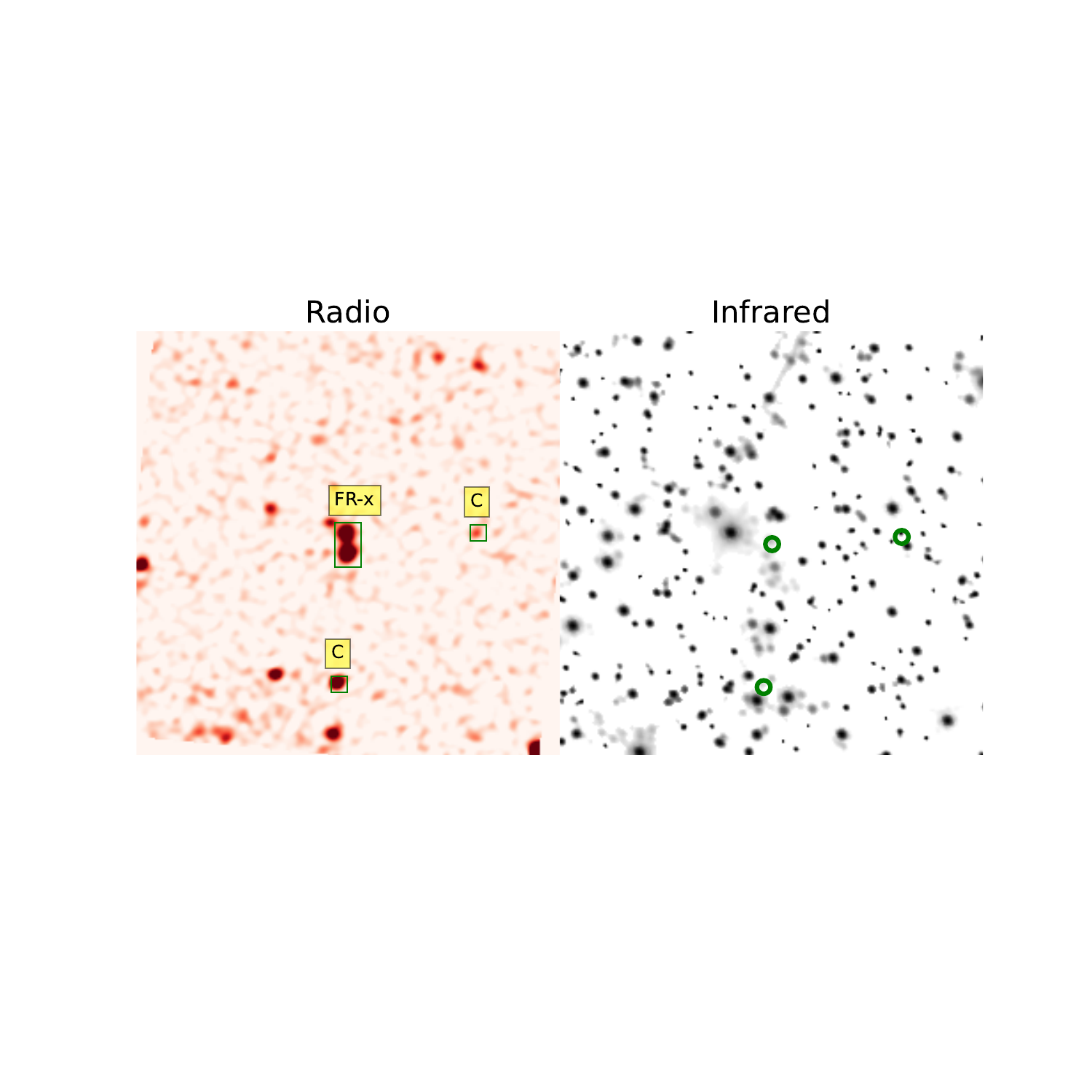}
\includegraphics[trim=2.9cm 3cm 2.9cm 12.5cm, width=8.8cm, scale=0.5]{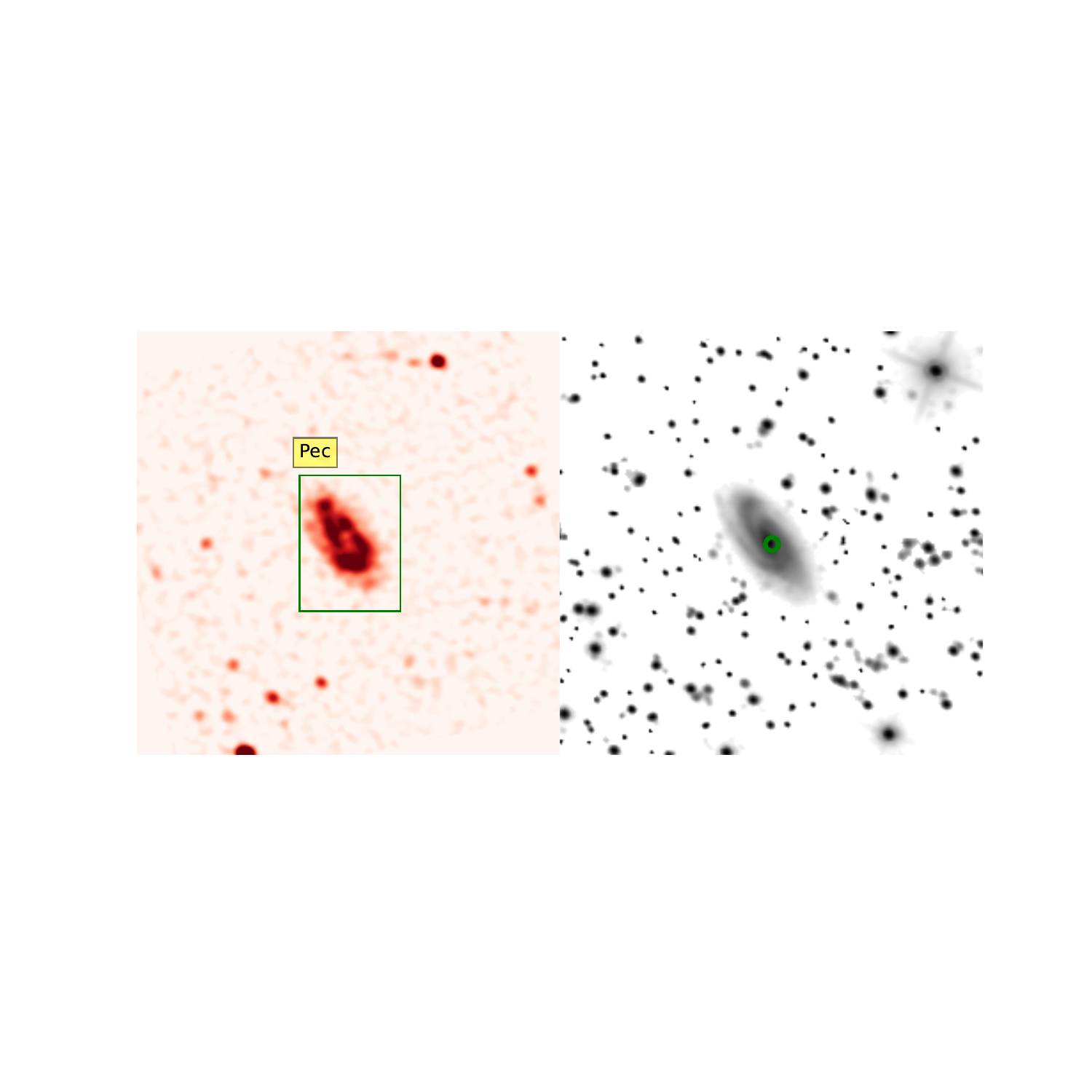}
\includegraphics[trim=2.9cm 3cm 2.9cm 12.5cm, width=8.8cm, scale=0.5]{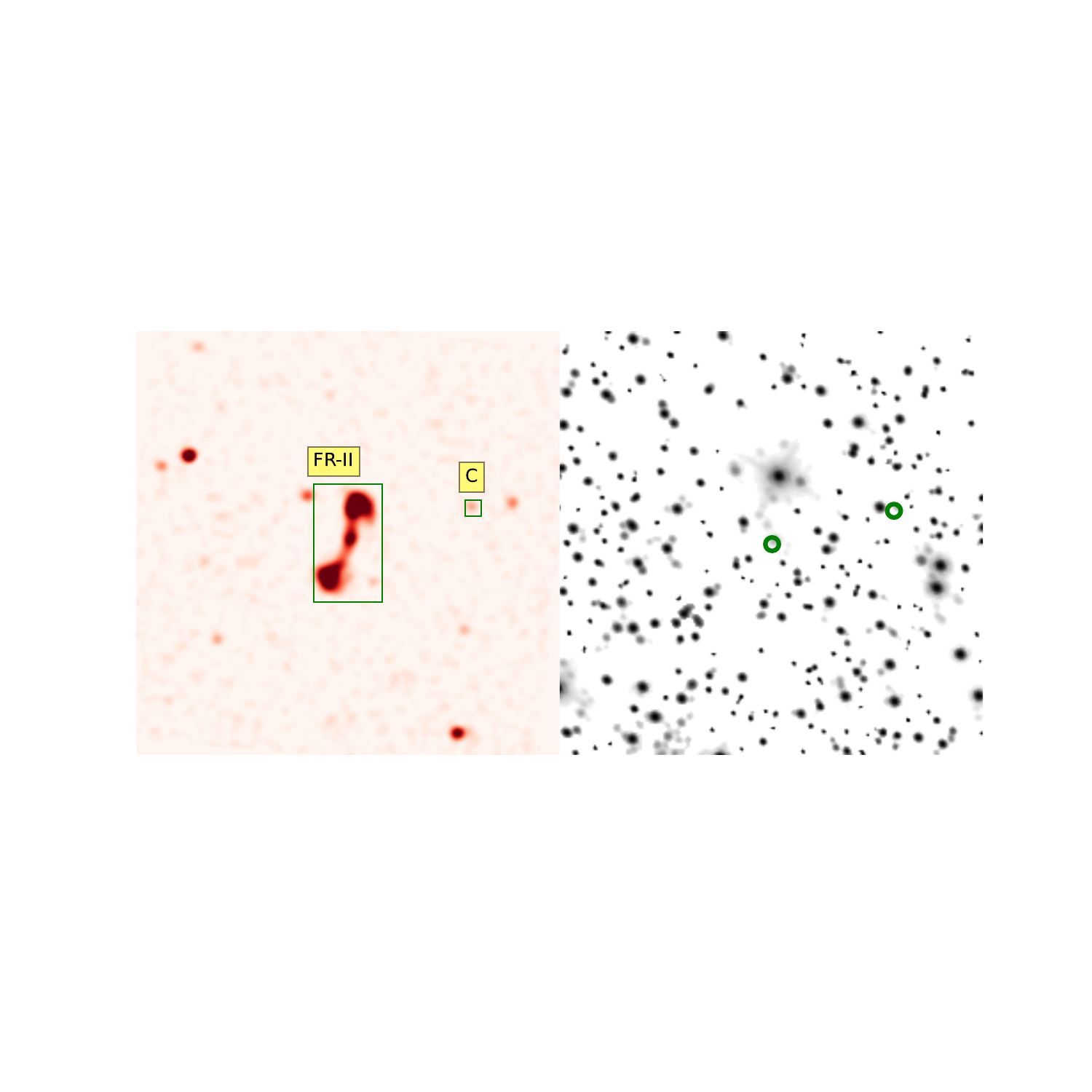}
\includegraphics[trim=2.9cm 3cm 2.9cm 12.5cm, width=8.8cm, scale=0.5]{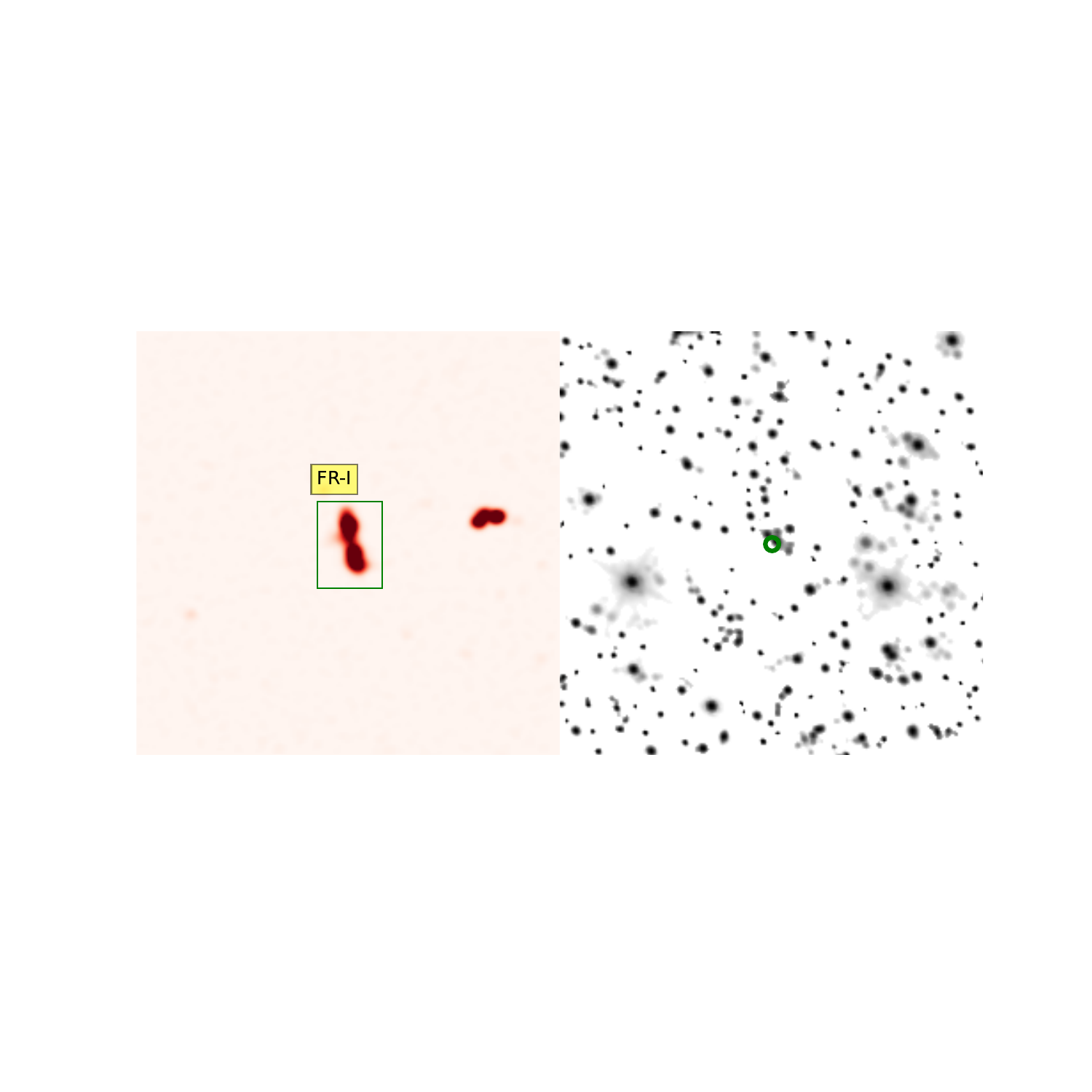}
\includegraphics[trim=2.9cm 3cm 2.9cm 12.5cm, width=8.8cm, scale=0.5]{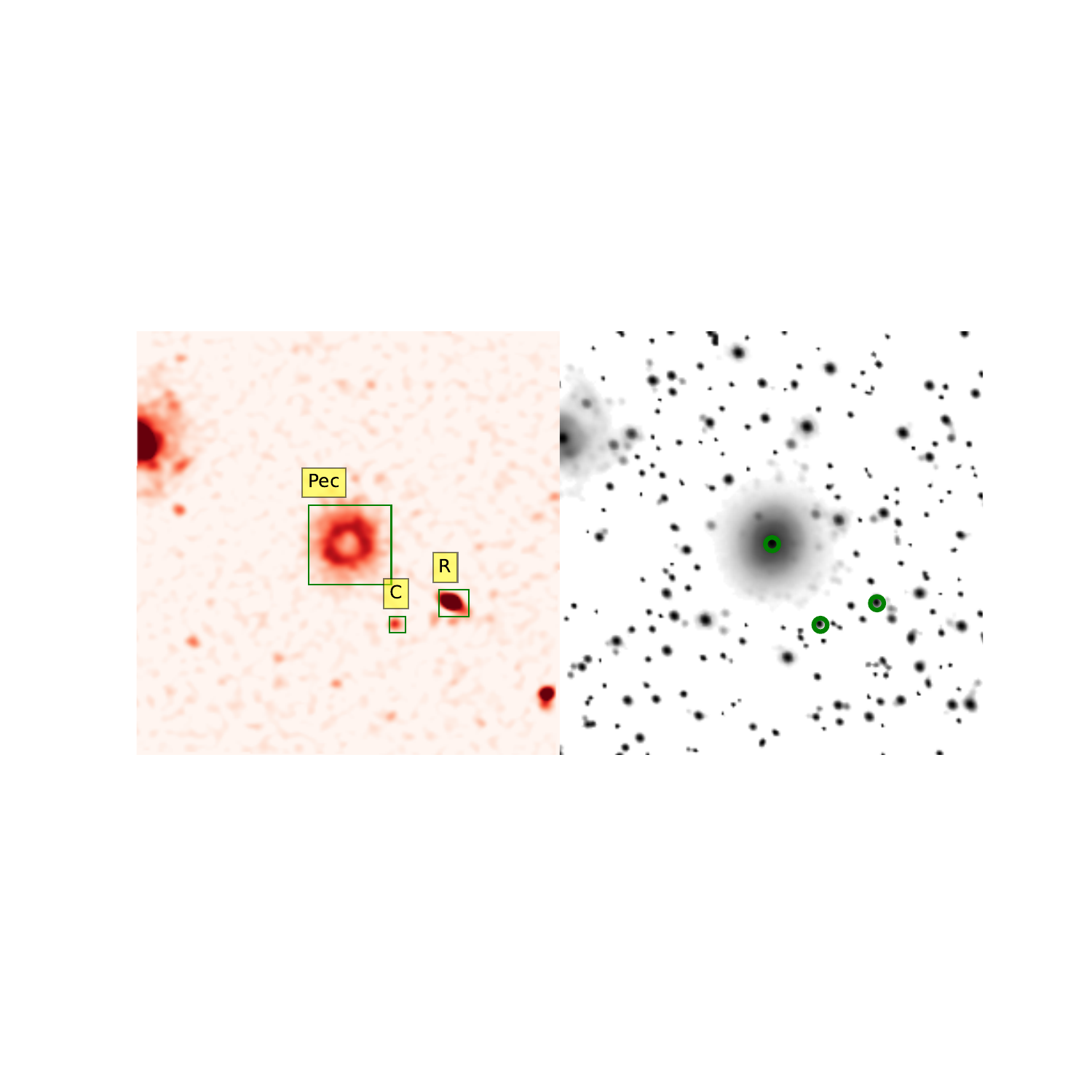}
\includegraphics[trim=2.9cm 3cm 2.9cm 12.5cm, width=8.8cm, scale=0.5]{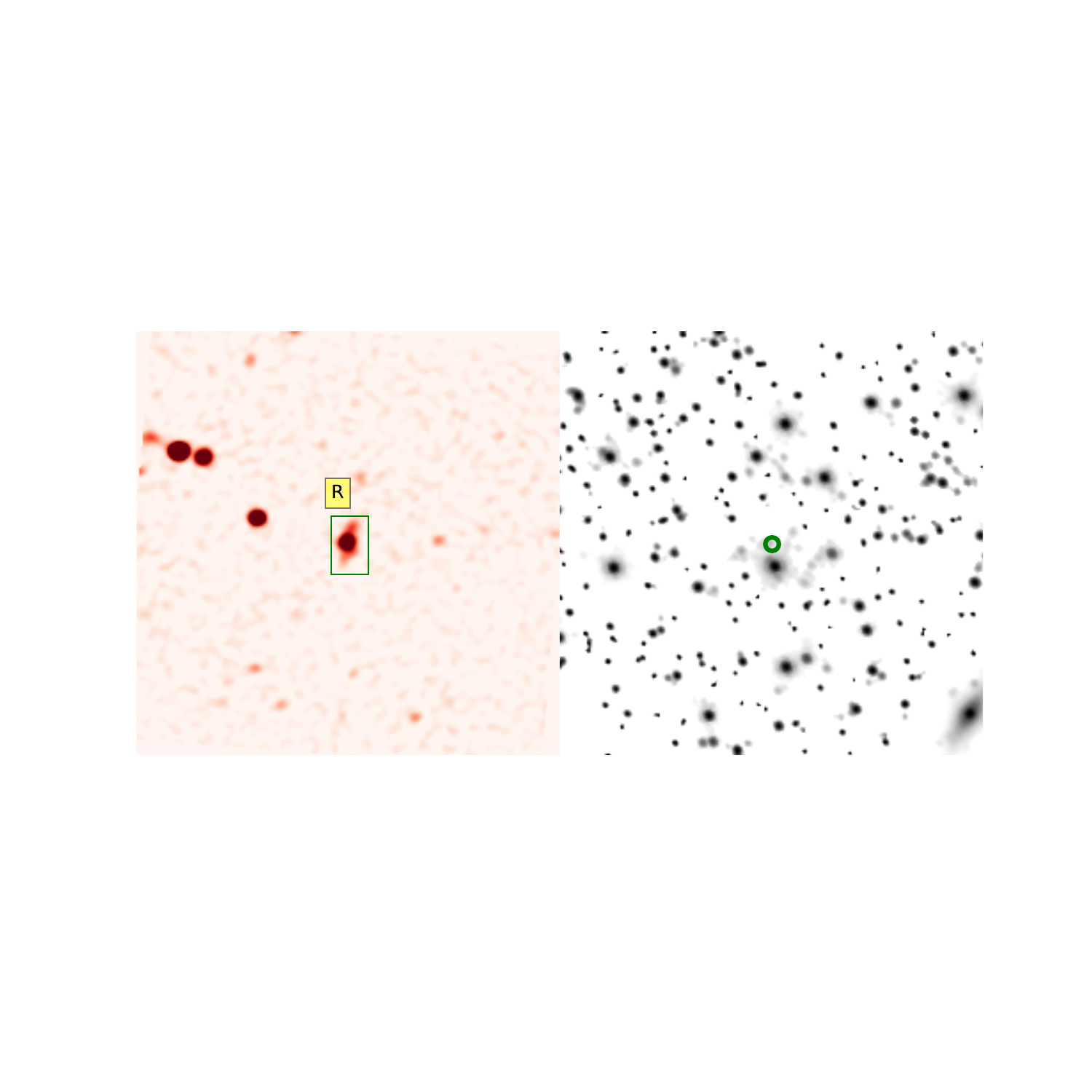}
\includegraphics[trim=2.9cm 3cm 2.9cm 12.5cm, width=8.8cm, scale=0.5]{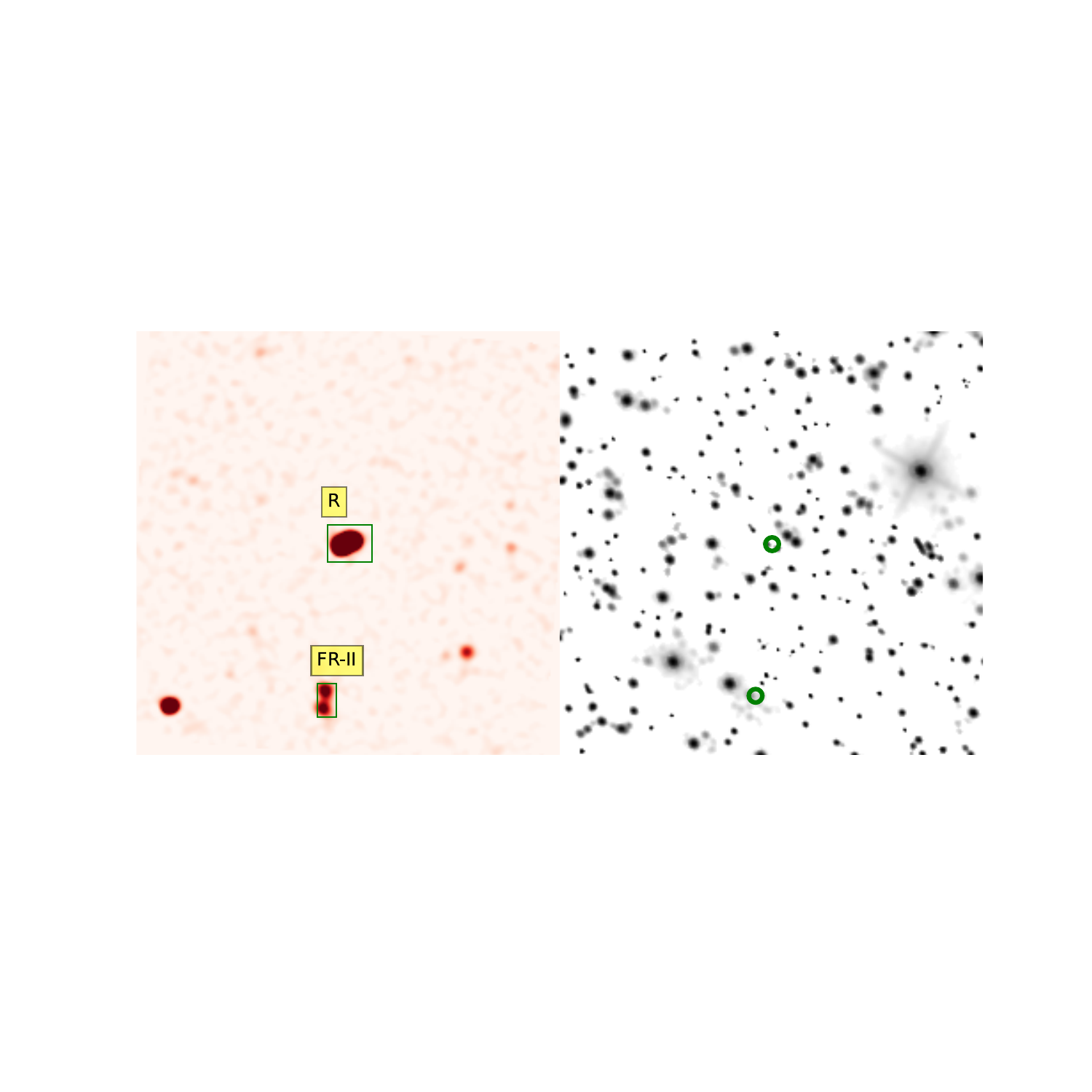}
\includegraphics[trim=2.9cm 8cm 2.9cm 12.5cm, width=8.8cm, scale=0.5]{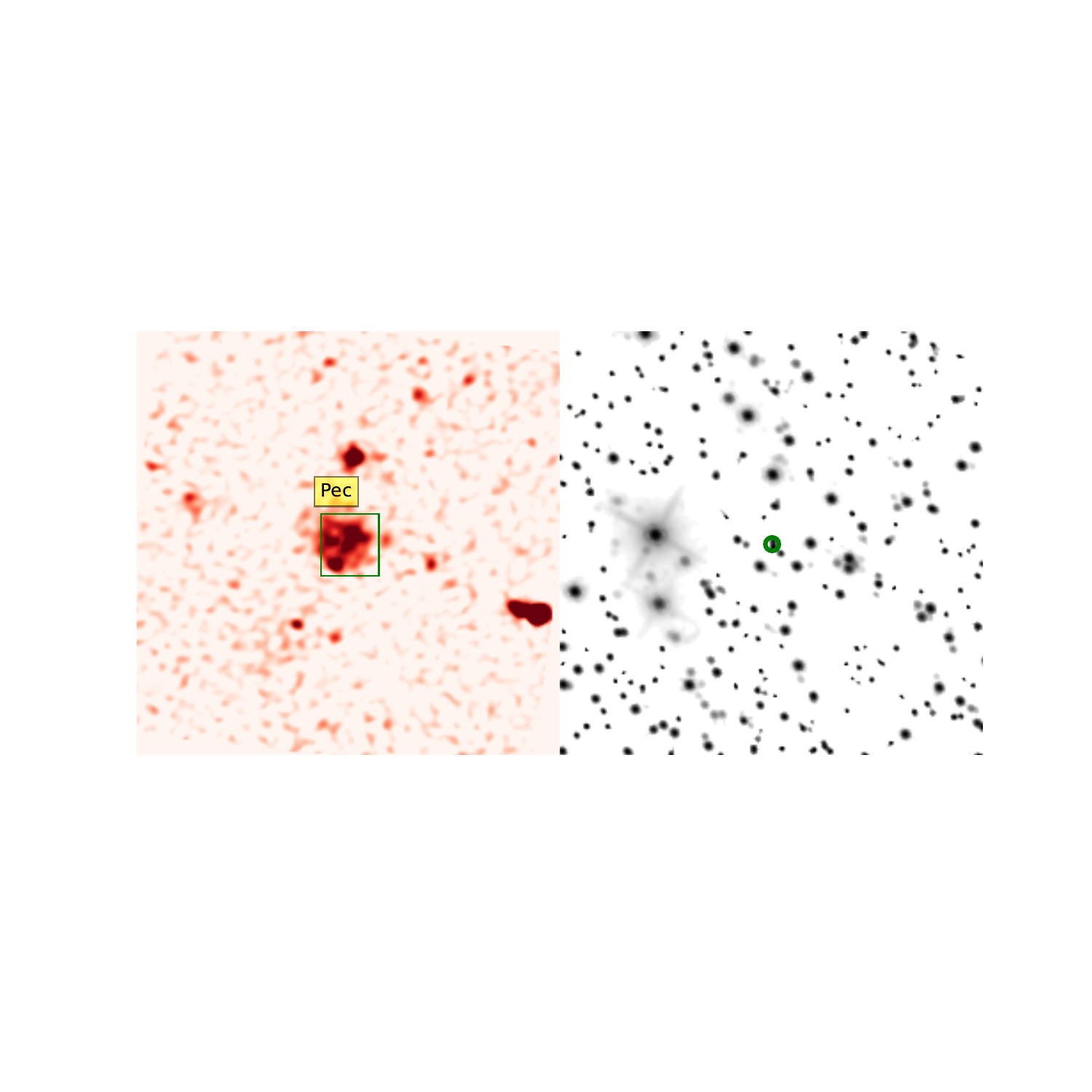}
\includegraphics[trim=2.9cm 8cm 2.9cm 12.5cm, width=8.8cm, scale=0.5]{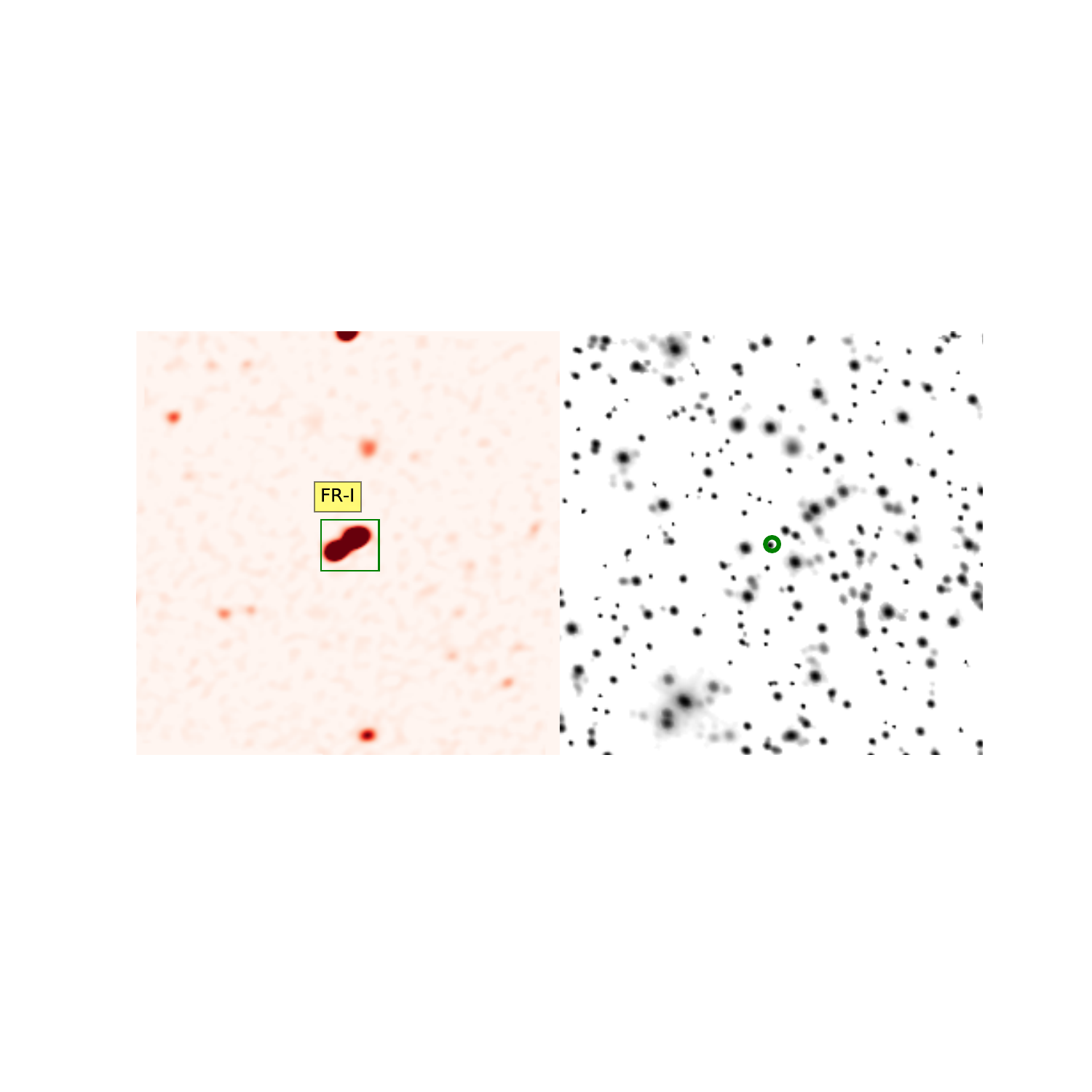}
\caption{Examples of the radio (left panels) and corresponding infrared (right panels) images, as described by the column titles. Each of these images has a frame size of $8^{\prime} \times 8^{\prime}$ in the sky ($240 \times 240$ pixels). In the radio images, we display classes and bounding boxes for radio galaxies encapsulating all their components. Here, the `FR-X' type is positioned between the FR-I and FR-II categories, 'R' denotes resolved radio sources with one visible peak, 'C' represents compact unresolved radio sources, and 'Pec' refers to peculiar or other rare radio morphologies (see Section~\ref{SEC:RadioIrAnnotations} for details). On the infrared images, circles indicate the positions of host galaxies.} 
\label{FIG:RadioGalaxies8arcm}
\end{figure*}

\section{Data}
\label{SEC:dataset}
In this section, we describe the radio and infrared observations, as well as the annotations developed and used for training the computer vision model to construct a consolidated catalogue.

\subsection{ASKAP Observations}
\label{SEC:ASKAP}
ASKAP, situated at Inyarrimahnha Ilgari Bundara, the Murchison Radio-astronomy Observatory (MRO), is a radio telescope equipped with PAF technology, enabling high survey speed through its wide instantaneous field of view. 
Comprising 36 antennas with various baselines, the majority are concentrated within a 2.3 km diameter region, while the outer six extend the baselines up to 6.4 km \citep{hotan21}. 
Recently, ASKAP concluded the first all-sky Rapid ASKAP Continuum Survey (RACS) \citep{McConnell20}, covering the entire sky south of Declination $+41^{\circ}$ with a median RMS of about 250 $\mu$Jy/beam. 
This has paved the way for subsequent deeper surveys using ASKAP.

The Evolutionary Map of the Universe (EMU), designed to observe the entire Southern Sky and potentially catalogue up to 40 million radio sources, is underway \citep[EMU;][]{norris11}. 
A significant step in this direction was the completion of the first EMU Pilot Survey \citep[EMU-PS][]{norris21} in late 2019. Covering 270 square degrees of the sky within $301^{\circ}< {\rm RA} < 336^{\circ}$ and $-63^{\circ}< {\rm Dec} < -48^{\circ}$, EMU-PS employed 10 tiles, each observed for approximately 10 hours. 
Achieving an RMS sensitivity between $25-35~\mu$Jy/beam and a beamwidth of $13^{\prime\prime} \times 11^{\prime\prime}$ FWHM, the survey operated in the frequency range of 800 to 1088 MHz, centred at 944 MHz.

The raw data from EMU-PS underwent processing using the ASKAPsoft pipeline \citep[][]{whiting17,norris21}. 
Since the survey comprised ten overlapping tiles, additional steps were taken for value-added processing to create a unified image and source catalogue. This involved merging the tiles by performing a weighted average of overlapping data regions and convolving the unified image to a standardized restoring beam size of $18^{\prime\prime}$ FWHM to address variations in the point spread function (PSF) from beam to beam \citep[][]{norris21}. 
The creation of a catalogue of islands and components was accomplished using the \textit{Selavy} source finder \citep{whiting12} applied to the convolved image, resulting in a compilation of 198,216 islands with 220,102 components, of which 90.3\% are single component islands and the rest are multiple component islands.
Note that the \textit{Selavy} catalogues are not designed to provide radio galaxy catalogues.
\textit{Selavy} initially detects pixels above a certain threshold, then groups those pixels into islands. Subsequently, components are fitted to each island, resulting in the creation of the component catalogue.
Note that throughout this work, we use unconvolved images with native resolution for tiles. The convolved image was only employed to generate the \textit{Selavy} catalogue \citep[see][for details]{norris21}.

\subsection{Infrared Observations}
\label{SEC:WISE}
For the EMU-PS, we obtain infrared images from the AllWISE observations of the Wide-field Infrared Survey Explorer (WISE) \citep[][]{wright10, cutri13}.
WISE conducted an all-sky infrared survey in the W1, W2, W3, and W4 bands, corresponding to wavelengths of 3.4, 4.6, 12, and 22 $\mu$m. 
Our study focuses on utilizing the W1 band from AllWISE, which has a 5$\sigma$ point source detection limit of 28 $\mu$Jy and angular resolution of $8.5^{\prime \prime}$.
In addition to images, we utilize the CatWISE catalogue \citep[][]{marocco21} of infrared sources to cross-match with the predicted infrared positions for constructing the consolidated catalogue.
The CatWISE catalogue is constructed from unWISE coadds \citep{lang14}, which have an angular resolution of $6.1^{\prime \prime}$.
The CatWISE Catalogue surpasses AllWISE by using six times as many exposures, resulting in approximately twice as many sources. 
Furthermore, CatWISE exhibits enhanced precision, especially for faint sources, resulting in a 12-fold improvement over AllWISE.

\subsection{Radio and Infrared Images}
\label{SEC:RadioInfraredImages}
The RadioGalaxyNET dataset \citep[][]{gupta2023b} comprises 2,800 extended radio galaxies and their infrared host galaxies identified in the EMU-PS and CatWISE through independent visual inspections, as detailed in \cite{gupta2023b} and \cite{yew22prep}. 
In addition to these extended radio galaxies, we incorporate compact ones into the dataset. 
Compact radio galaxies refer to unresolved radio sources that do not show extended emission.
We begin by initially selecting sources randomly from the  \textit{Selavy} catalogue, wherein the algorithm identifies only one component corresponding to an island. 
Subsequently, we utilize the CARTA visualization tool \citep{comrie21} to validate their compact nature and identify their respective infrared hosts in the EMU-PS and AllWISE images, respectively. 
These infrared hosts are determined as the nearest galaxies to the radio peaks identified during visual inspections. 

A visual inspection of 2,700 single-component radio sources in \textit{Selavy} reveals that 146 of them are associated with extended radio sources, suggesting that approximately 95\% of these single-component radio sources identified by \textit{Selavy} are genuinely compact.
Furthermore, we refine this dataset of compact radio sources by crossmatching it with the CatWISE catalogue to confirm the visual identifications of infrared hosts. 
In line with the methodology proposed by \citet{norris21}, we employ a maximum cross-matching separation of $3^{\prime \prime}$ between visually identified infrared positions and CatWISE positions to ensure that the false identification rate remains below 8\%.
Our ultimate dataset of compact radio galaxies includes 2,090 verified infrared host galaxies. 
In the present study, we exclusively utilize these compact radio galaxies to train and test the deep-learning computer vision model, excluding compact sources with unconfirmed infrared hosts. 
It is important to note that the \textit{Selavy} catalogue of EMU-PS includes approximately 180,000 sources with single components. 
However, a small randomly selected subset is included here to maintain the balance between other categories. 

In addition to the compact radio galaxies, we also introduce peculiar and other rare radio morphologies into the dataset. 
Instances of these morphologies, as illustrated in \citet{gupta22}, encompass peculiar-shaped extended emission, diffuse emission from galaxy clusters, very nearby, thus large and resolved star-forming galaxies, and peculiars like Odd Radio Circles \citep[ORCs;][]{norris21b}. 
Initially, we seek out these sources in the EMU-PS image to identify any similar morphologies.
Subsequently, our exploration extends to identify other morphologies that significantly deviate from the common types. 
Examples of these include oddly bent extended radio galaxies and neighbouring extended radio galaxies with merging emissions, presenting challenges in classification within conventional radio morphologies. 
Our investigation yields a total of 99 such peculiar and other rare morphologies, all integrated into the training and evaluation dataset.
Note that not all of these morphologies are radio galaxies; some of them form a group with multiple hosts. 
Therefore, for simplicity, we utilize the terms `galaxy' or `morphology' interchangeably for all these sources.
Additionally, while some of these morphologies, such as ORCs, fall under the category of peculiar sources, others like resolved star-forming, oddly bent galaxies etc. are rare but not considered peculiar. However, for brevity, we use the abbreviation `Pec' for all of these 100 morphologies.
With a composition of 2,800 extended radio galaxies in RadioGalaxyNET, along with 2,090 compact radio galaxies and 100 sources with peculiar and other rare morphologies, our dataset comprises approximately 5,000 radio galaxies, forming a comprehensive set for training and testing our object detection deep learning model.

\begin{figure*}[!t]
\centering
\includegraphics[width=17.5cm, scale=0.5]{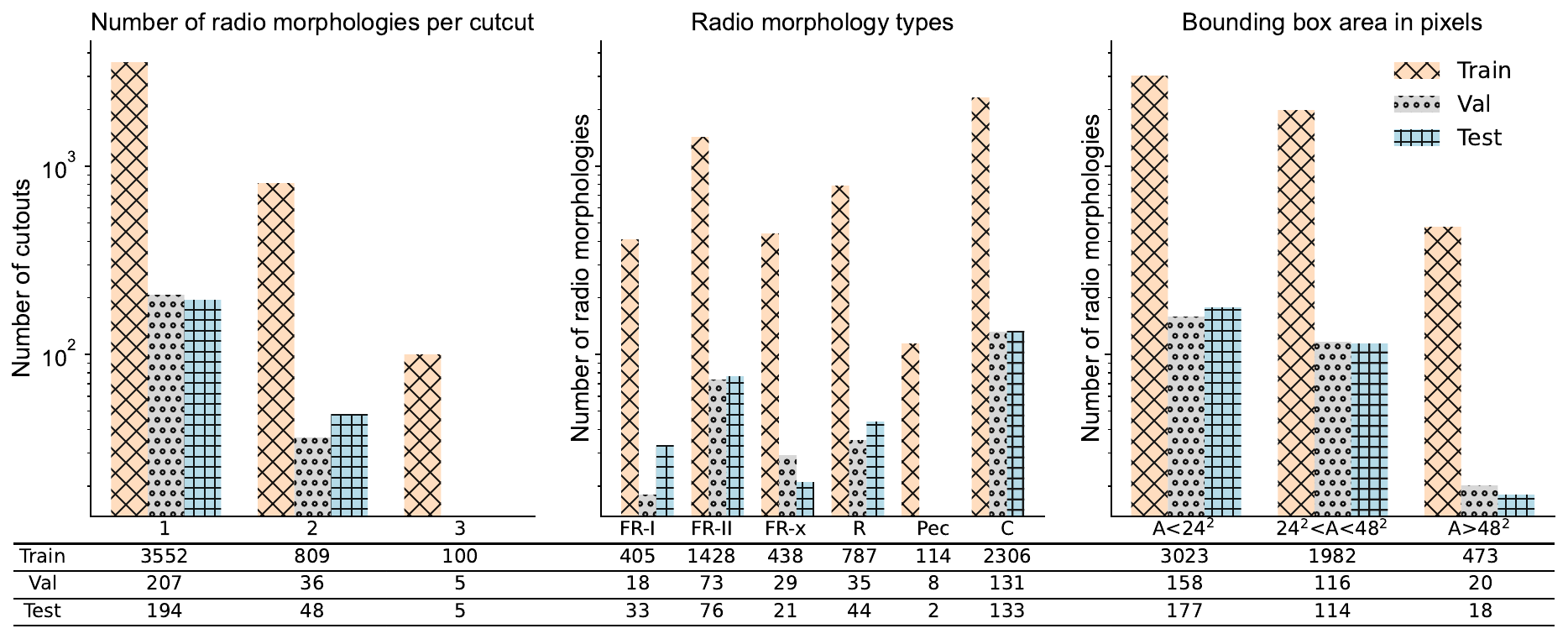}
\caption{Shown are the dataset split distributions, depicting the distributions of extended radio galaxies in a single cutout (left), their respective categories (middle), and the occupied area per radio galaxy (A; right). The tables below the figures provide detailed counts of radio morphologies in the training, validation, and test sets. See Section~\ref{SEC:Datastats} for more details. Note that each radio morphology has a corresponding infrared host, so the counts here also represent the number of corresponding infrared hosts.}
\label{FIG:Statistics}
\end{figure*}

We create image cutouts with dimensions of $8^{\prime} \times 8^{\prime}$ in the sky, resulting in images of $240 \times 240$ pixels, where each pixel corresponds to 2$^{\prime \prime} \times$ 2$^{\prime \prime}$. 
Examples of these radio images are illustrated in the left panels of Figure~\ref{FIG:RadioGalaxies8arcm}. 
It is important to note that the cutouts of radio images in RadioGalaxyNET are initially $15^{\prime} \times 15^{\prime}$, designed to detect several distinct instances of these multi-component radio galaxies in the same image. However, for our current objective of grouping multiple components of central galaxies to construct the catalogue, we reduce the cutout size, as detailed in Section~\ref{SEC:Catalog}.
This is motivated by the fact that only 10 out of the 5,000 extended radio galaxies in our dataset have a total extension larger than $7^{\prime}$.
Future work should incorporate these very large extended radio galaxies into the dataset as we identify more such galaxies in the ongoing main EMU survey. 
However, for our current study, we have not included them in the training.
In contrast to RadioGalaxyNET images, we refrain from preprocessing our radio maps in this study. Instead, the network is adapted to handle FITS files for radio images, as explained in Section~\ref{SEC:MLMethods}.
At the identical sky locations of radio images, we acquire infrared images from AllWISE, sourced from the WISE W1 band.
We create cutouts of the same size as the radio images and then reproject the infrared images onto the radio images using the world coordinate system.
In contrast to radio images, infrared images undergo noise reduction processing using the method detailed in \cite{gupta2023a, gupta2023b}. Examples of these processed infrared images are showcased in the right columns of Figure\ref{FIG:RadioGalaxies8arcm}.

\subsection{Radio and Infrared Annotations}
\label{SEC:RadioIrAnnotations}
Annotations for 2,800 extended radio galaxies and their corresponding infrared hosts are already present in the RadioGalaxyNET dataset \citep[][]{gupta2023b}. 
These annotations encompass radio galaxy categories, bounding boxes encapsulating all radio components of each radio galaxy, pixel-level segmentation masks, and the keypoint locations of corresponding infrared hosts that are cross-matched with the CatWISE catalogue. 
The identification of these extended radio galaxies and their infrared hosts is independently achieved through visual inspections, where the infrared host for each radio galaxy is identified in the infrared image \citep{yew22prep}.
The dataset is categorized into FR-I, FR-II, FR-x, and R galaxies based on measurements of their total extent and the distance between peak positions. 
This classification follows the criteria set by \cite{fanaroff74}, where the ratio between the peak distance and total extent is employed to differentiate between FR-I and FR-II galaxies. 
Specifically, the ratio is below 0.45 for FR-I and above 0.55 for FR-II. 
Owing to image resolution limitations, some galaxies cannot be conclusively classified as either FR-I or FR-II, leading to their categorization as FR-x galaxies, with the ratio between peak distance and total extent falling between 0.45 and 0.55. 
Note that, in this work, certain subtypes of morphologies, which are commonly associated with FR-I type radio galaxies, such as bent-tailed, narrow, and wide-angled-tailed sources \citep{miley80}, are not differentiated.
The R category (radio galaxies with resolved radio emission) pertains to those galaxies where only one central peak is visible, resulting in a ratio set to zero.

For compact radio galaxies, denoted by `C', we initially apply the island segmentation method \citep[e.g.,][]{gupta22} to mask pixels exceeding $3\sigma$ for compact radio galaxies. 
Subsequently, we generate bounding boxes for these galaxies. 
The positions of infrared hosts for these compact radio galaxies are identified through visual inspections and included in the dataset as keypoints.
Note that R and C radio galaxies used for model training and evaluation are visually distinguished in this work and do not rely on the peak-to-total flux criterion \citep[e.g.,][]{condon98}.
In the case of peculiar and other rare radio morphologies, denoted as `Pec', we employ CARTA to obtain bounding boxes and subsequently generate segmentation masks for all pixels larger than $3\sigma$ within these bounding boxes. 
Due to the diffuse nature of radio emission in these Pec morphologies, the hosts for these sources cannot be identified unambiguously with a unique infrared object, sometimes involving multiple potential host galaxies. 
Consequently, we select one of these infrared galaxies, closest to the bounding box centroid, as the host for the Pec radio morphologies. 
Future work, incorporating a larger sample of Pec radio morphologies, should include all such infrared galaxies to enable multiple host galaxy detection for these sources.
Figure~\ref{FIG:RadioGalaxies8arcm} provides examples of FR-I, FR-II, FR-x, R, and C radio galaxies and Pec morphologies.
It is worth noting that the FR-I radio galaxies depicted in the middle left and bottom right panels may indeed be FR-II-type radio galaxies at higher resolution. Nevertheless, we adhere to the mechanism described above and classify them based on the present image resolution.
Following the RadioGalaxyNET dataset structure, we furnish annotations for the radio images, encompassing `categories,' `bbox,' and `segmentation,' along with `keypoints' for the infrared. All annotations adhere to the COCO dataset format \citep{lin14M}, simplifying the streamlined evaluation of object detection methods.

\subsection{Radio Source Statistics}
\label{SEC:Datastats}
Our dataset encompasses approximately 5,000 radio galaxies and their corresponding infrared hosts. 
We generate 5,000 cutouts centered at each radio galaxy, each with a size of $8^{\prime} \times 8^{\prime}$, and store them in FITS image format with sky information in the headers. 
Similarly, for infrared host galaxies, we generate cutouts of the same size and process them following the methodology outlined in Section~\ref{SEC:RadioInfraredImages}. 
As other radio galaxies are present close to the central radio galaxy within an $8^{\prime} \times 8^{\prime}$ cutout (e.g., top right panel of Figure~\ref{FIG:RadioGalaxies8arcm}), there are a total of 6,080 instances of radio galaxies in these 5,000 cutouts.
The dataset comprises 371 FR-I, 1,331 FR-II, 401 FR-x, 698 R, and 2,090 compact radio galaxies as well as 99 other rare and peculiar morphologies. 
Adhering to the commonly used strategy in machine learning, we randomly split our dataset into train, validation, and test sets in the ratio of $0.9:0.05:0.05$ for the object detection modelling. 

The specific counts of radio galaxies within each radio-morphological category, along with the split ratios, are illustrated in Figure~\ref{FIG:Statistics} and detailed in the table below the figure.
The left panel of the figure showcases the number of $8^{\prime} \times 8^{\prime}$ cutouts with radio galaxy instances ranging from single to three radio galaxies, indicating that the majority of our cutouts feature either single or double instances of radio galaxies. 
Note that while these cutouts with multiple instances are utilized for training and testing the network, catalogue construction is based solely on central radio galaxies and their infrared host galaxies, as discussed in Section~\ref{SEC:Catalog}. 
The middle panel of the figure presents the number of radio galaxies within the six categories, with the corresponding counts for the split sets displayed in the table below. 
Lastly, the third panel reveals the number of radio galaxies categorized by bounding box area in pixels, highlighting that the majority of our radio galaxies are small-scale structures, with bounding box area below $48^2$ pixels.

\section{Detection Pipeline for Radio Galaxies}
\label{SEC:MLMethods}
The radio images exhibit distinct features for extended radio galaxies compared to their counterparts in infrared images, where the latter predominantly resemble point sources (with a few exceptions, such as resolved star-forming galaxies).
Examples of these disparities are illustrated in Figure~\ref{FIG:RadioGalaxies8arcm}. 
A multimodal approach, introduced by \citet{gupta2023b}, facilitates the detection of both radio sources and their potential infrared host positions. 
This approach incorporates models like Gal-DETR, Gal-Deformable DETR, and Gal-DINO, all capable of concurrently identifying radio galaxies and their potential infrared hosts.
These models employ two fundamental detection schemes: the base networks handle class and bounding box predictions for radio galaxies, while the keypoint detection module is utilized for predicting potential infrared host positions.
Table 2 in \cite{gupta2023b} demonstrates that the Gal-DINO\footnote{\url{https://github.com/Nikhel1/Gal-DINO}} model, outperforms other networks in our context of small object detection in radio and infrared images.
For a comprehensive understanding of the modelling strategy, we direct readers to \cite{gupta2023b}; here, we provide a brief overview of the Gal-DINO model.

The Gal-DINO model is based on the DEtection TRansformers (DETR) \citep{carion2020end}, which adopts the Transformer architecture \citep{vaswani2017attention}. 
Initially designed for natural language processing, this architecture is utilized to address the intricate task of object detection in images. 
In contrast to conventional methods relying on region proposal networks \citep[e.g., Faster RCNN;][]{ren2015faster}, DETR introduces an end-to-end approach to object detection using Transformers.
The DETR with Improved deNoising anchOr boxes \citep[DINO;][]{zhang2023dino} incorporates enhanced anchor boxes, predefined boxes crucial for object detection. 
DINO introduces refined strategies for selecting and placing these anchor boxes, improving the model's capability to detect objects of varying sizes and aspect ratios. 
During training, DINO employs an improved mechanism for matching anchor boxes to ground truth objects, enhancing accuracy in localization and classification. 
Additionally, DINO utilizes adaptive convolutional features, enabling the model to concentrate on informative regions of the image, thereby enhancing both efficiency and accuracy.
Gal-DINO \citep{gupta2023b} integrates keypoint detection into the DINO algorithm, which already features improved de-noising anchor boxes. 
By mitigating the impact of noise and outliers, Gal-DINO yields more resilient and precise bounding box predictions. 
This results in improved localization of extended radio galaxies and their corresponding infrared hosts within these bounding boxes.

\begin{figure}[!t]
\centering
\includegraphics[width=7.5cm, scale=0.5]{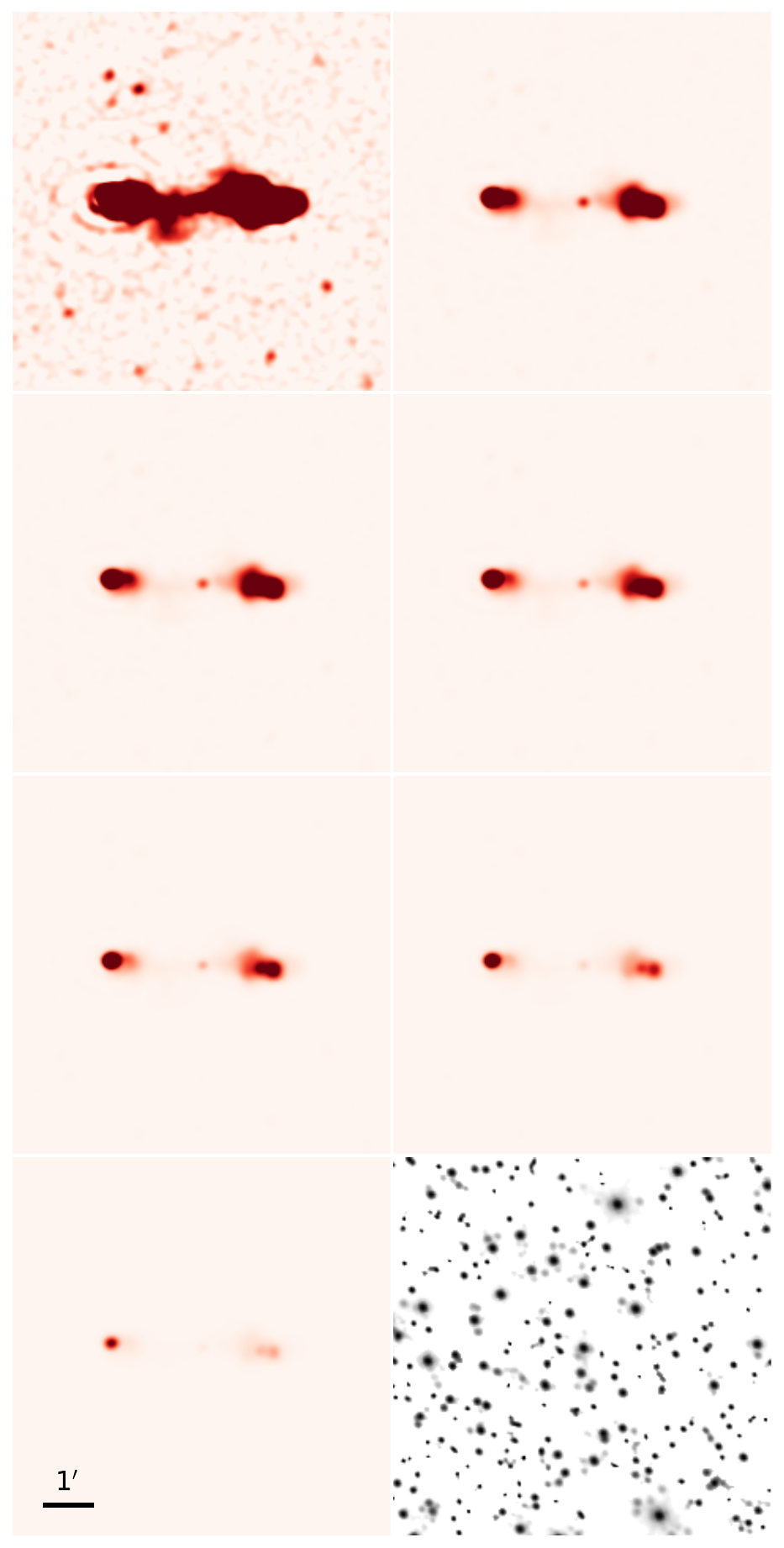}
\caption{Shown is an example of an 8-channel image used for the training and evaluation of the Gal-DINO network. The first 7 channels contain data from the radio FITS file, representing the extended radio galaxy with clipping between the 50th percentile level and 7 specific maxima, corresponding to the 95th, 99th, 99.2th, 99.5th, 99.7th, 99.9th, and 99.99th percentile levels. The 8th channel, in the bottom right, displays the corresponding pre-processed infrared image. The bounding box and keypoint annotations are not depicted here for brevity; examples of these annotations are shown in Figure~\ref{FIG:RadioGalaxies8arcm} on the radio (with maxima at the 95th percentile level) and infrared images, respectively.}
\label{FIG:processing}
\end{figure}

\subsection{Radio Galaxy Class and Bounding Box Predictions}
The Gal-DINO model utilizes the DINO architecture, integrating a ResNet-50 \citep{he15d} Convolutional Neural Network backbone to process input images and extract feature maps. 
Positional encodings are introduced to incorporate spatial information into the Transformer architecture, enhancing the model's ability to understand relative object positions. 
DINO uses learned object queries, departing from fixed anchor boxes \citep[e.g.][]{tian2019fcos}, to represent classes of objects and are refined during training. 
The model's decoder produces two output heads for class and bounding box predictions, facilitating simultaneous processing of the entire image and capturing contextual relationships between regions. 
Each head refers to a specific sub-network for a particular aspect of the overall task.
This approach results in robust and precise bounding box predictions, improving the localization of extended radio galaxies.
DINO adopts the Hungarian loss function to establish associations between predicted and ground-truth bounding boxes, ensuring a one-to-one mapping. 
The comprehensive loss function for DINO integrates the cross-entropy loss for class predictions $\mathcal{L}_{\text{c}}$ and the smooth L1 loss for bounding box predictions $\mathcal{L}_{\text{b}}$.
This loss function is expressed as the sum of the classification and bounding box losses, contributing to the overall training objective of the model.
\BE
\mathcal{L}_{\text{DINO}} = \mathcal{L}_{\text{c}} +  \mathcal{L}_{\text{b}}.
\label{EQ:LossDETR}
\EE
The L1 loss, used for bounding box regression, measures the absolute difference between predicted and true bounding box coordinates, penalizing the model for deviations and promoting accurate localization.

\subsection{Infrared Host Galaxy Keypoint Prediction}
Gal-DINO introduces keypoint detection techniques as a complement to the bounding box-based object detection method for identifying extended radio galaxies and their potential infrared hosts. 
Keypoints, representing distinctive features within images, offer precise spatial information for accurately locating the host galaxy. 
Unlike bounding boxes, keypoints allow for fine-grained localization, particularly beneficial for complex radio emission morphologies. 
The keypoint detection in Gal-DINO, leveraging the transformer-based architecture to capture global and local details. 
It utilizes self-attention mechanisms to localize and associate keypoints for infrared host galaxies. 
The loss function for Gal-DINO combines DINO loss for class and bounding box predictions with keypoint detection loss, expressed as
\BE
\mathcal{L}_{\text{Gal-DINO}} = \mathcal{L}_{\text{DINO}} +  \mathcal{L}_{\text{k}}
\label{EQ:LossGalDETR}
\EE
where $\mathcal{L}_{\text{k}}$ is the L1 loss for keypoint detection, calculating the absolute difference between predicted and ground truth keypoint coordinates, such as the x and y position of the host.

\subsection{Network Training}
\label{SEC:Training}
The dataset is split into training, validation, and test sets as illustrated in Figure~\ref{FIG:Statistics} and explained in Section~\ref{SEC:Datastats}. 
The training set is utilized to train the networks, while the validation and test sets function as inference datasets during and after training, respectively. 
As mentioned in Section~\ref{SEC:RadioInfraredImages}, the Gal-DINO network employed in the present work, processes radio images in FITS format and infrared images in PNG format, diverging from the RadioGalaxyNET dataset where 3-channel radio-radio-infrared PNG format files are used for both training and inference.
In the present approach, 8-channel images are generated, comprising 7 channels from the radio FITS file and 1 channel representing the processed infrared sky. 
Each radio channel encompasses clipped data from FITS files, where clipping is done between the 50th percentile level and 7 distinct maximum levels.
These maxima correspond to the 95th, 99th, 99.2nd, 99.5th, 99.7th, 99.9th, and 99.99th percentile levels. 
This methodology utilizes radio images in FITS format directly, eliminating the need for preprocessing and conversion to PNG files. Furthermore, it imparts information to the computer vision model in a manner akin to the visual inspection performed by expert astronomers for the classification and grouping of multi-component radio galaxies.
Future work should explore alternative scaling approaches and, if necessary, fine-tune the number of channels while taking into account potential GPU memory constraints.
An illustrative example of an FR-II radio galaxy, featuring 7 channels and its corresponding infrared channel, is presented in Figure~\ref{FIG:processing}.

In line with the training methodology of Gal-DINO in \cite{gupta2023b}, the training dataset undergoes diverse random augmentations during each epoch, where an epoch corresponds to a single pass through the entire dataset in the model's training process. 
These augmentations encompass horizontal flipping, random rotations (-180 to 180 degrees), random resizing within $400\times400$ to $1300\times1300$ pixels, and random cropping of a randomly selected set of 8-channel training images. 
Horizontal flipping and rotations ensure varied orientations, crucial for handling radio galaxies with diverse sky orientations while resizing and cropping introduce scale and spatial variety. 
These augmentations, consistently applied, bolster the model's ability to generalize effectively and excel on unseen data. 
Gal-DINO, with a parameter count of 47 million, underwent a training duration of approximately 35 hours using a single Nvidia Tesla P100 GPU for 100 epochs.
We utilise the original hyperparameters from Gal-DINO with specific adjustments. 
These hyperparameters, set before training, encompass critical aspects such as learning rate, batch size, architecture details, dropout rate, activation functions, and optimizer. 
Although a comprehensive list of hyperparameters for each network is not provided here for brevity, detailed network architecture, including hyperparameters, is available in the associated repositories for reference.

\subsection{Evaluation Metrics}
\label{SEC:Evaluate}
The evaluation metrics, based on \cite{coco14}, employ the Intersection over Union (IoU) to assess algorithm performance on the test dataset. 
IoU is calculated as the ratio of the overlap area between predicted ($B_{\rm P}$) and ground truth ($B_{\rm GT}$) bounding boxes to their union area. 
\begin{equation}
    \text{IoU}(B_{\rm P}|B_{\rm GT}) = \frac{{\text{Overlap between $B_{\rm P}$ and $B_{\rm GT}$}}}{{\text{Union between $B_{\rm P}$ and $B_{\rm GT}$}}},
\end{equation}
In bounding box prediction, each predicted box is categorized as true positive (TP), false positive (FP), or false negative (FN) based on its area compared to the ground truth box. A TP bounding box accurately identifies an object with high IoU overlap, while an FP bounding box fails to correspond to any ground truth object. An FN bounding box occurs when an object in the ground truth data is not successfully detected by the algorithm, representing a missed opportunity to identify a genuine object.
To detect keypoints, we use the Object Keypoint Similarity (OKS) metric, which gauges the similarity between predicted and ground-truth keypoints. 
OKS computes the Euclidean distance for each predicted keypoint relative to its corresponding ground-truth keypoint, normalizing it based on the size of the object instance.
The Euclidean distance ($Ed$) between the ground truth and predicted keypoints is subjected to a Gaussian function, defined as follows:
\begin{equation}
\mathrm{OKS} = \exp \left(-\frac{Ed^2}{2l^2c^2}\right),
\end{equation}
where $l$ represents the ratio of the bounding box's area to the image cutout area, and $c$ is a keypoint constant set to 0.107. 
This adjustment differs from the 10 values used in \cite{coco14}, as each bounding box in our case corresponds to a single infrared host. The resulting OKS score falls within the range of 0 to 1, with 1 signifying perfect keypoint localization.

\begin{table}[!ht]
    \centering
    \caption{Results for bounding box and keypoint detection using the trained Gal-DINO network are presented on a combination of the test and validation datasets with 8-channel images (see Figure~\ref{FIG:processing}). The columns, from left to right, showcase various metric types: average precision for IoU (or OKS) thresholds ranging from 0.50 to 0.95 (AP), a specific IoU (or OKS) threshold of 0.5 (AP$_{50}$), IoU (or OKS) threshold of 0.75 (AP$_{75}$), and average precision for small-sized (AP$_{\rm S}$), medium-sized (AP$_{\rm M}$), and large-sized (AP$_{\rm L}$) radio galaxies. Further details on the training and evaluation can be found in Sections~\ref{SEC:Training} and \ref{SEC:Evaluate}, respectively.}
    \begin{NiceTabular}{lcccccc}
    \toprule
      Metric    & AP  & AP$_{50}$ & AP$_{75}$  & AP$_{\rm S}$ & AP$_{\rm M}$ & AP$_{\rm L}$ \\
      Type      & (\%)  & (\%)  & (\%)  & (\%) & (\%) & (\%) \\
    \midrule
     Bbox     & 65.1  & 73.2  & 70.0  & 69.5  &  65.4 & 66.9 \\
     Keypoint & 69.3  & 71.7  & 70.0  & 55.5  &  93.8 & 97.2 \\
    \bottomrule
    \end{NiceTabular}
    \label{TAB:AP1}
\end{table}

We evaluate the network's performance using the average precision metric, a widely used standard in object detection model assessment \citep[][]{coco14}. 
Precision, the ratio of true positives to the total number of objects identified as positive, measures detection and classification precision. 
Recall, the ratio of true positives to the total number of objects with ground truth labels, assesses the model's ability to identify all relevant objects. 
The precision-recall curve illustrates the trade-off between precision and recall across varying detection thresholds. The area under the curve (AUC) is calculated, representing the average AUC value across all classes or objects. 
Higher values of average precision, which ranges from 0 to 1, signify superior model performance. 
We compute average precision using standard IoU and OKS thresholds for bounding boxes and keypoints. 
IoU and OKS thresholds determine the correctness of predictions based on overlap and similarity scores, respectively. 
We calculate average precision at IoU (or OKS) thresholds from 0.50 to 0.95 (AP), as well as specific thresholds of 0.50 (AP$_{50}$) and 0.75 (AP$_{75}$) for radio galaxies of all sizes. 
Additionally, we assess performance across different structure scales by computing Average Precision for small (AP$_{\rm S}$), medium (AP$_{\rm M}$), and large (AP$_{\rm L}$) area ranges defined by pixel areas $\rm A < 24^2$, $\rm 24^2 < A < 48^2$, and $\rm A > 48^2$, as shown in the right panel of Figure~\ref{FIG:Statistics}.

\subsection{Model Evaluation Results}
\label{SEC:MLResults}
We assess the performance of the Gal-DINO model using 8-channel radio and infrared images that were not part of the training dataset. 
The model is designed to predict the categories (FR-I, FR-II, FR-x, R, Pec, and C) of these galaxies, generate bounding boxes to capture their extended emission structures, and identify their corresponding infrared host galaxies. 
Details about the radio and infrared images, annotations, and data statistics can be found in Sections~\ref{SEC:RadioInfraredImages}, \ref{SEC:RadioIrAnnotations}, and \ref{SEC:Datastats}, respectively. 
The model's training and evaluation strategy is explained in Sections~\ref{SEC:Training} and \ref{SEC:Evaluate}, respectively. 
Evaluation is performed on a combined dataset of validation and test sets. 
This combined evaluation is justified by the need for larger sample sizes; for example, the test set only contains 2 Pec morphologies, but combining the validation set increases this count to 10. 
It is important to note that the validation set plays no role in training the network; it is solely utilized for evaluating the model at each epoch without model back-propagation applied using the epoch's validation results. 
Thus, it is appropriate to merge both the validation and test sets for the evaluation of the trained model.

\begin{figure}[!t]
\centering
\includegraphics[width=7.5cm, scale=0.5]{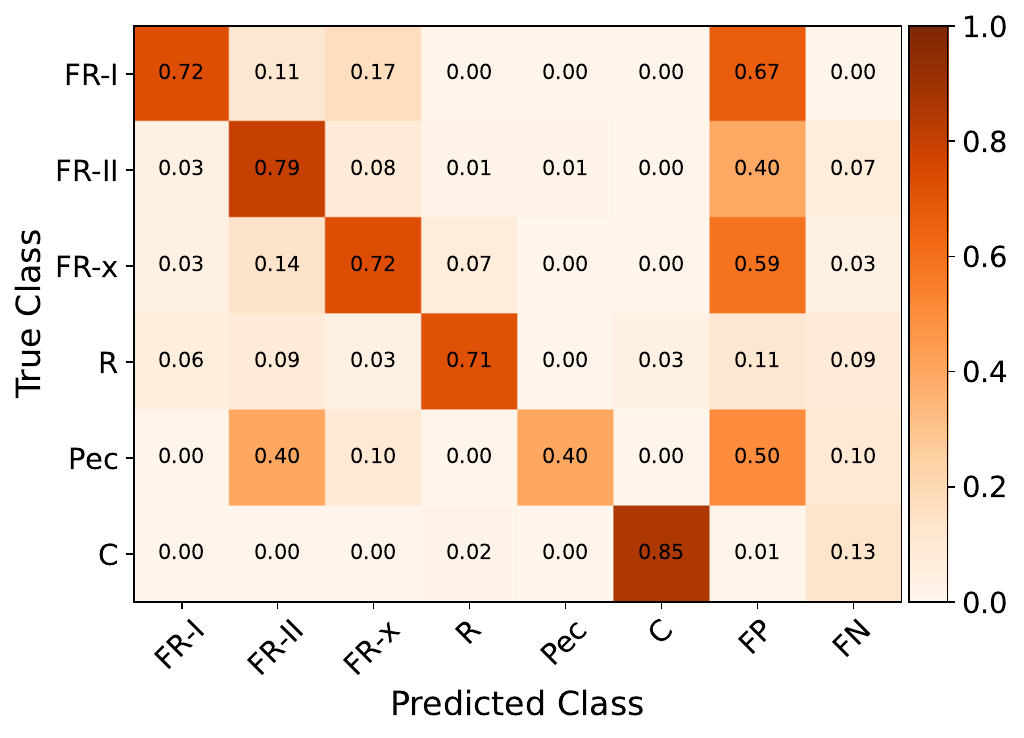}
\caption{Presented is the normalized confusion matrix for the Gal-DINO detection model. Each matrix is normalized based on the total number of galaxies within its corresponding class. The diagonal entries denote true positive (TP) instances, representing objects correctly detected with an IoU and OKS threshold surpassing 0.5 compared to the ground truth instances, and a confidence threshold of 0.25. False positive (FP) instances correspond to model detections lacking corresponding ground truth instances, while false negative (FN) instances signify objects that the model failed to detect at the same IoU and OKS thresholds, along with a confidence threshold of 0.25.} 
\label{FIG:Confusion}
\end{figure}

Table~\ref{TAB:AP1} presents the outcomes obtained from the combined validation and test dataset for both bounding box and keypoint detections. 
The Gal-DINO model, after training, attains an AP of 65.1\% for bounding boxes associated with radio galaxies. 
Its effectiveness extends across various IoU thresholds, achieving an AP$_{50}$ of 73.2\% and an AP$_{75}$ of 70.0\%, demonstrating robustness in detecting radio galaxies at different IoU thresholds. 
Furthermore, the model consistently performs well for radio galaxies of small, medium, and large sizes (AP$_{\rm S}$, AP$_{\rm M}$, and AP${\rm L}$). 
In keypoint detection as well, our trained model demonstrates strong performance, attaining an AP of 69.3\% and an AP$_{50}$ of 71.7\%. 
These results substantiate the model's efficacy in predicting bounding boxes and keypoints for both radio galaxies and their corresponding infrared hosts, showcasing strong performance across diverse IoU thresholds and bounding box areas.

\begin{table}[!ht]
    \centering
    \caption{The bounding box and keypoint detection results achieved through the Gal-DINO network on a merged dataset comprising both test and validation sets. The columns correspond to those outlined in Table~\ref{TAB:AP1}. The PNG results reflect outcomes obtained from 3-channel images, while the All-10\% results signify a scenario where 10\% of the entire training dataset is intentionally corrupted, aiming to assess the model's robustness to potential noisy labels. The Ext-10\% results specifically involve introducing 10\% noise in annotations exclusively for extended radio galaxies within the training dataset (see Section~\ref{SEC:MLMethods} for details).}
    \begin{NiceTabular}{llcccccc}
    \toprule
     & Metric    & AP  & AP$_{50}$ & AP$_{75}$  & AP$_{\rm S}$ & AP$_{\rm M}$ & AP$_{\rm L}$ \\
     & Type      & (\%)  & (\%)  & (\%)  & (\%) & (\%) & (\%) \\
    \midrule
    \Block{2-1}{\rotate PNG}
     &  Bbox     & 63.8  & 69.4  & 69.3  & 68.5  &  58.9 & 45.0 \\
     &  Keypoint & 65.6  & 67.5  & 66.5  & 51.5  &  93.6 & 97.0 \\
    \midrule
    \Block{4-1}{\rotate All-10\%}
     &  Bbox     & 63.4  & 71.9  & 68.7  & 66.8  &  65.4 & 65.7 \\
     &  (std)    & $\pm2.5$ & $\pm3.1$  & $\pm2.5$  & $\pm3.1$  &  $\pm4.2$ & $\pm14.8$ \\
     &  Keypoint & 67.7  & 70.5  & 68.5  & 54.2  &  93.9 & 96.7 \\
     &  (std)    & $\pm2.1$  & $\pm2.0$  & $\pm2.2$  & $\pm2.3$  &  $\pm4.4$ & $\pm1.9$ \\
    \midrule
    \Block{4-1}{\rotate Ext-10\%}
     &  Bbox     & 64.9  & 73.0  & 70.1  & 68.2  &  62.2 & 66.6 \\
     &  (std)    & $\pm2.1$ & $\pm3.4$  & $\pm2.8$  & $\pm3.3$  &  $\pm3.8$ & $\pm15.5$ \\
     &  Keypoint & 68.7  & 71.5  & 69.1  & 53.7  &  89.9 & 96.4 \\
     &  (std)    & $\pm2.3$  & $\pm1.6$  & $\pm2.4$  & $\pm2.4$  &  $\pm4.1$ & $\pm2.4$ \\
    \bottomrule
    \end{NiceTabular}
    \label{TAB:AP2}
\end{table}

Figure~\ref{FIG:Confusion} presents the confusion matrices for the combined validation and test set, computed based on IoU for bounding boxes (and OKS for keypoints) at confidence thresholds of 0.5 and 0.25, respectively. 
It is crucial to highlight the distinction between confusion matrices used in object detection and classification. 
Object detection involves the possibility of multiple instances of the same or different classes within an image, resulting in several true positive (TP), false positive (FP), and false negative (FN) values for each class. 
In contrast, classification usually assumes only one label per image, yielding a confusion matrix with a single TP, FP, and FN value for each image.

Precision in object detection depends on precise localization and accurate detection of object boundaries, evaluated using an IoU (and OKS) threshold of 0.5. 
The confusion matrix contains TP, FP, and FN values, offering insights into the model's performance. 
For the FR-II class, the model achieved a TP value of 0.79, indicating correct detection of 79\% of FR-II galaxy instances. 
However, there were significant false positives (FP = 0.40), where the model predicted 40\% of instances as FR-II when they did not correspond to FR-II instances in the ground truth. 
Additionally, a moderate number of false negatives (FN = 0.07) suggests that the model missed or failed to detect 7\% of actual FR-II instances. Similar patterns are observed for the FR-I, FR-x, and Pec classes.
For the FR-I class, the detected source extent depends on sensitivity to diffuse extended structure \citep[e.g., see Figure 14 in][]{turner18}, and hence, some FR-I sources may be classified as FR-IIs due to limited surface brightness sensitivity.

Peculiar and other rare morphologies exhibit around 40\% misclassification as FR-II galaxies, primarily due to similarities in some rare morphologies to FR galaxies but with unique diffuse emission or wide-angled-tailed characteristics. 
It is important to highlight that although this percentage may seem high, it corresponds to only four Pec morphologies.
Compact radio galaxies achieve a high TP rate of 85\%, with a few misclassified as R galaxies due to only slight differences. 
Although the false negative rates in the detections are not optimal, it is important to emphasize that these result from the application of a confidence threshold of 0.25.
Lowering this threshold, eliminates FN rates for all categories, indicating the detection of all radio galaxies in the ground truth. 
Nonetheless, this leads to higher FP rates, a topic that will be revisited in Section~\ref{SEC:Catalog} for catalogue construction.

As outlined in Section~\ref{SEC:RadioInfraredImages}, the images utilized in \cite{gupta2023b} for both training and model evaluation consist of 3-channel PNG images with dimensions of $450\times450$ pixels ($15^{\prime}\times 15^{\prime}$). 
These images involve two pre-processed radio channels, containing 0-8 and 8-16 bit information, and one corresponding pre-processed infrared channel, with 8-16 bit information.
In this work, we diverge from this approach by training the network with 8-channel images sized at $240\times240$ pixels ($8^{\prime}\times 8^{\prime}$), as illustrated in Figure~\ref{FIG:processing} and detailed in Section~\ref{SEC:Training}. 
In Table 2 of \cite{gupta2023b}, it is evident that the AP$_{50}$ for bounding boxes is 60.2\%, whereas in our current work, the AP$_{50}$ attains 73.2\%, as demonstrated in Table~\ref{TAB:AP1}.
To understand this difference, we conduct an experiment where the Gal-DINO model is trained using our dataset (Section~\ref{SEC:Datastats}) and the same set of hyperparameters for 100 epochs. 
However, in this case, the network is trained with 3-channel PNG images containing radio-radio-infrared channels, mirroring the approach in \cite{gupta2023b}.
The evaluation results using these PNG images are presented in Table~\ref{TAB:AP2}. 
When considering the combined validation and test sets, the AP$_{50}$ is measured as 69.4\% for bounding boxes.
This comparison shows that employing 8-channel images in our current work marginally improves model performance in comparison to the utilization of 3-channel PNG images, as indicated by the evaluation results. 
Additionally, it reaffirms the anticipated outcome that reducing image dimensions from $450\times450$ pixels to $240\times240$ pixels significantly contributes to improved model performance when compared with the findings in \cite{gupta2023b}. 

\begin{figure*}[!ht]
\centering
\includegraphics[width=18.cm, scale=0.5]{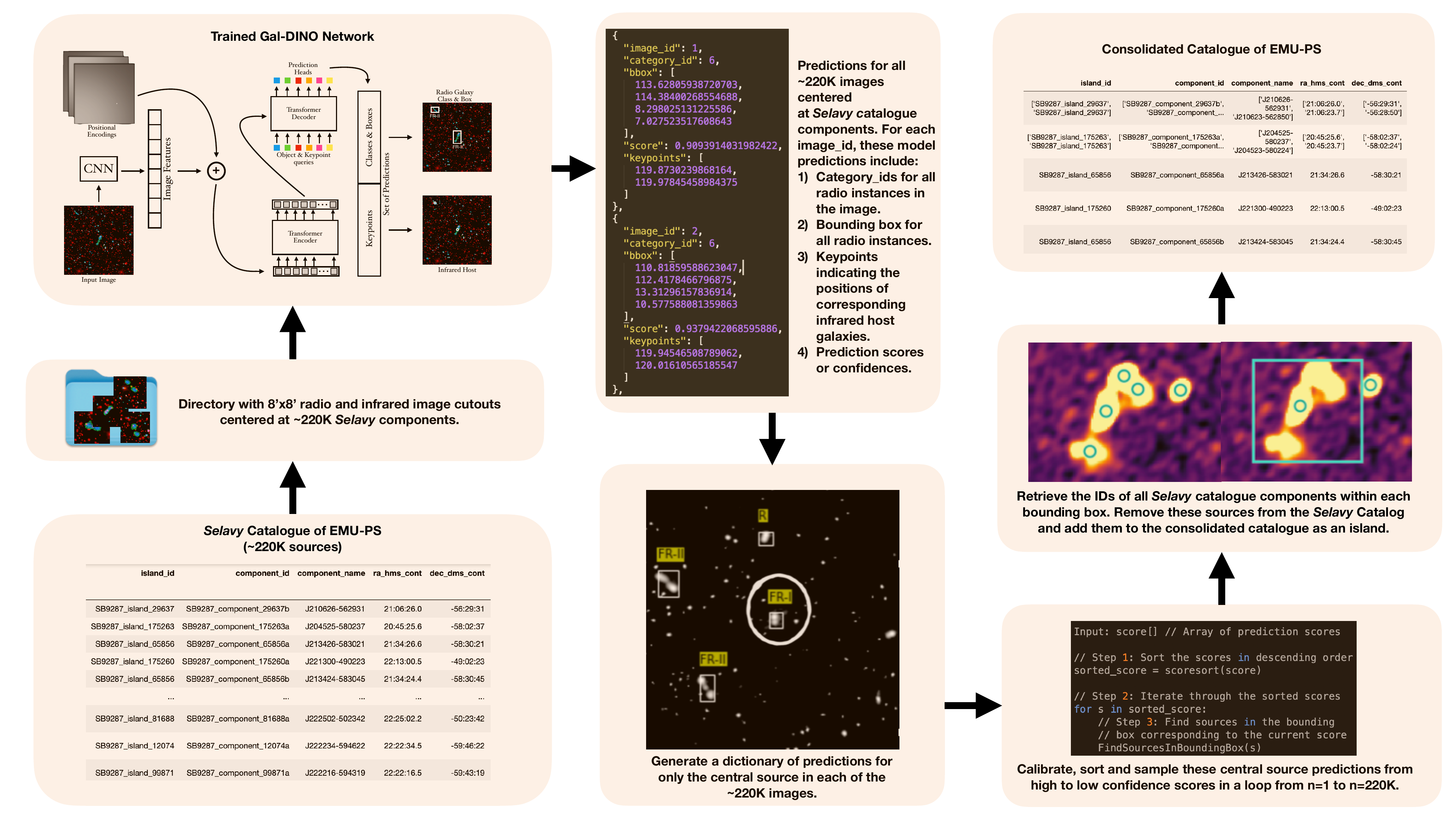}
\caption{An overview of the catalogue construction pipeline. The process initiates with obtaining predictions from the Gal-DINO model for all radio and infrared cutouts centred at the components in the \textit{Selavy} catalogue. Subsequently, a dictionary of predictions is generated for the central sources within these cutouts. The consolidated catalogue is then formed by calibrating the confidence scores in the dictionary, organizing them in descending order, and systematically consolidating and removing entries from the \textit{Selavy} catalogue based on decreasing score values. We refer the readers to Section~\ref{SEC:Catalog} for further details.} 
\label{FIG:CAT_Overview}
\end{figure*}

Given that visual inspections are employed for annotating the dataset used in training the network, we assess the model's robustness to possible errors introduced during the manual labelling process. 
To achieve this, we undertake an experiment wherein the model is trained on noisy labels that are intentionally corrupted with noise.
In this experiment, we introduce noise by randomly altering the bounding boxes for 10\% of the training set. 
Specifically, we modify the bounding box positions by displacing them away from the full extent of the radio galaxy. 
This ensures that these noisy bounding boxes and keypoints neither coincide with the galaxy nor maintain a consistent placement in the image; instead, they are randomly positioned elsewhere within the image. 
This process is repeated ten times, each time selecting different galaxies at random. 
The model is trained in the same manner and for the same number of epochs as before, with five distinct training datasets, each with 10\% randomly selected galaxies carrying noisy labels.
Note that this noise is only applied to the training dataset, keeping the validation and test datasets unchanged for a direct comparison with our main results outlined in Table~\ref{TAB:AP1}. 
Table~\ref{TAB:AP2} presents the outcomes when the entire training set is employed to randomize the bounding boxes, denoted as All-10\%.
The median AP$_{50}$ for the five models trained with random noisy data at a 10\% level is $71.9\pm 3.1$ for bounding box detections. 
Here, the error is computed as the standard deviation across the results of five model evaluations on the validation and test datasets.
This observation suggests that the model remains unaffected even when 10\% of the annotations in the training dataset are erroneous. 
Comparable results are observed for AP, AP$_{75}$, AP$_{S}$, AP$_{M}$, and AP$_{L}$ metrics. It's worth noting that the larger error bars in AP${L}$ may be attributed to the smaller validation and test sample size, as illustrated in the right panel of Figure~\ref{FIG:Statistics}.

In another analogous experiment, we again altered the positions of the bounding boxes and keypoints by randomly selecting 10\% of the extended radio galaxies. 
This selection encompasses all five categories of radio galaxies, excluding compact radio galaxies this time. 
This is conducted to assess the model's robustness to possible noise in annotations related to extended radio galaxies.
In this procedure, we randomly select 10\% of these extended radio galaxies and proceed to randomly displace their bounding boxes and keypoint positions, following the same methodology as before. 
Over five such iterations, the model is trained in a manner consistent with the previous experiment.
The outcomes for these models, denoted as Ext-10\%, utilizing noisy annotations for extended radio galaxies during model training, are presented in Table~\ref{TAB:AP2}. 
Notably, the results for both bounding boxes and keypoints consistently align with the primary evaluation findings reported in Table~\ref{TAB:AP1}. 
This demonstrates the model's robustness against potential manual errors in the annotation process, such as incorrect radio morphology, radio bounding boxes, and infrared host galaxies.

\section{Catalogue Construction Pipeline}
\label{SEC:Catalog}
The Gal-DINO model, once trained, is employed to generate a catalogue of radio galaxies within the EMU-PS.
Figure~\ref{FIG:CAT_Overview} provides an overview of this process. 
This section will delve into the specifics of the catalogue construction pipeline.
Note that, unless the existence of confirmed infrared or optical counterparts is established, these entities are generally referred to as radio sources rather than AGN or radio galaxies. 
However, as our pipeline identifies potential hosts for all radio sources, we use the term `radio galaxies' for brevity. 
The cross-matching process of counterpart identification is elaborated in Section~\ref{SEC:MultiwavelengthCounterparts}.

\subsection{Images for Catalogue Construction}
\label{SEC:SelavyImages}
We employ the \textit{Selavy} catalogue of the EMU-PS to extract cutouts, which, in conjunction with the trained Gal-DINO model, provide radio galaxy and its potential infrared host predictions. As outlined in Section~\ref{SEC:ASKAP}, the primary purpose of \textit{Selavy} catalogues is not to generate radio galaxy catalogues; rather, their design only focuses on grouping pixels into islands and fitting components to each island.
Utilizing the \textit{Selavy} source finder on the EMU-PS image resulted in identifying 220,102 components. We create cutouts, sized at $8^{\prime}\times8^{\prime}$, at the positions of these components, where radio cutouts are saved as FITS files, and corresponding pre-processed infrared cutouts are saved as PNG files. These cutouts are saved in a directory and then processed by the trained Gal-DINO model to obtain predictions for each cutout.
It is important to note that components in the \textit{Selavy} catalogue can appear in multiple images due to their proximity. Consequently, the Gal-DINO model may generate multiple predictions for each component across different cutouts. As detailed in Section~\ref{SEC:SelavyCentralSources}, only central radio source instances in each image are utilized for catalogue construction. Since each image has a unique \textit{Selavy} catalogue component at its centre, this allows us to make predictions for each source individually.

\begin{figure*}[!ht]
\centering
\includegraphics[width=17.cm, scale=0.5]{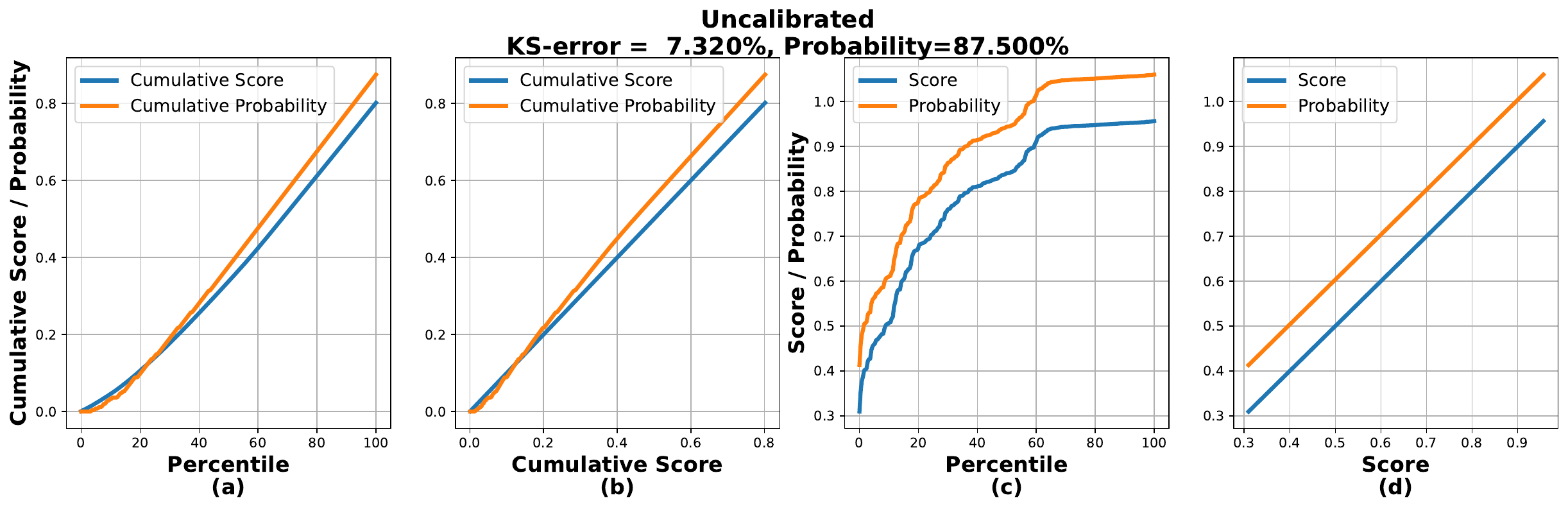}
\includegraphics[width=17.cm, scale=0.5]{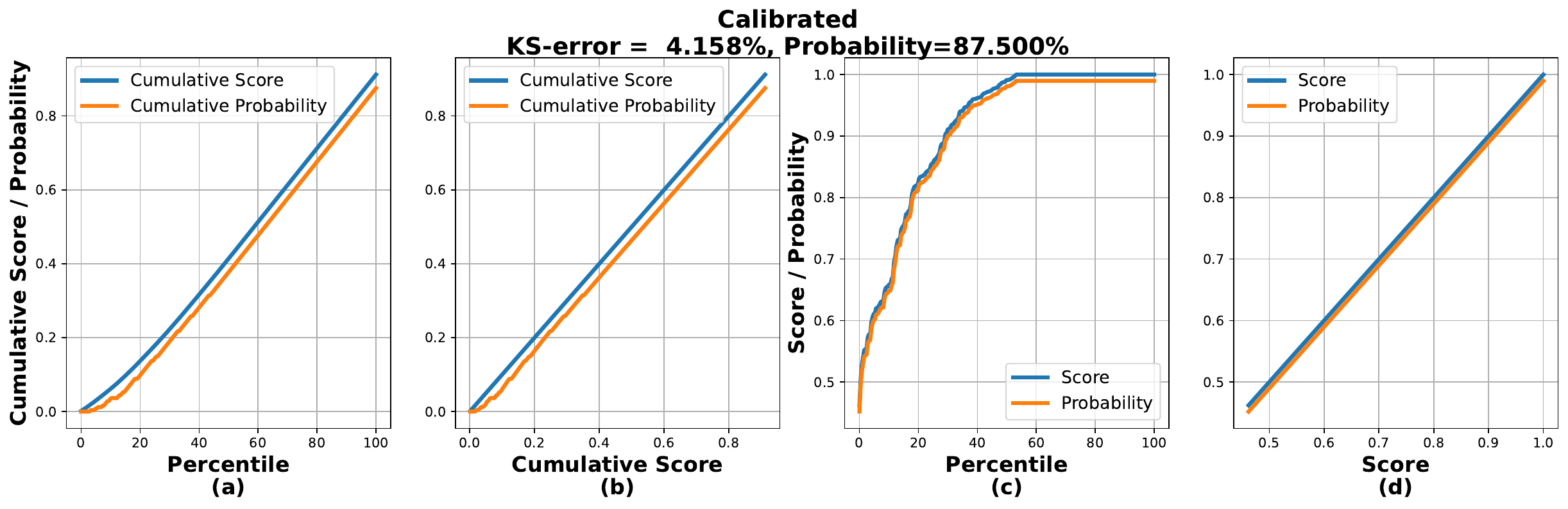}
\caption{Uncalibrated and calibrated scores for combined validation and test sets: (a) depicts the cumulative score and probability with the fractile. In (b), the same data is shown, but with a distorted horizontal axis, resulting in a cumulative score graph forming a straight line. This is visualized through scatter plots of cumulative (score, score) in blue and (score, probability) in orange. In the case of a perfectly calibrated network, the probability line will coincide as a straight line with the (score, score) line.
(c) and (d) showcase plots of non-cumulative scores and probabilities against the fractile or score. The upper panel demonstrates that the network significantly overestimates the probability of detection based on the score with a KS error of 7.3\%. The calibration reduces the KS error to 4.2\%. See Section~\ref{SEC:CataloguingWithScore} for more details.} 
\label{FIG:ScoreCalib}
\end{figure*}

\subsection{Model Predictions for Images}
\label{SEC:SelavyPredictions}
The directory containing 220K radio and infrared image cutouts is input into the trained Gal-DINO model through its data loader. 
The model then provides predictions for each detected radio source, including a category assignment, a bounding box encompassing any extended emission or multiple components, and a prediction score or confidence level. 
Additionally, the model predicts potential infrared hosts for all identified radio sources within the provided cutouts.
For 220K cutouts, the model requires approximately 10 hours for these predictions, when running on a single Nvidia Tesla P100 GPU.
While the network produces several FP predictions, many of these predictions possess very low detection scores.
To filter out less reliable predictions, we establish a detection score threshold, only retaining predictions with a score greater than 0.05. 
While this choice reduces the false positive detection rate, it also maintains the false negative rate at a minimal level. 
Employing this score threshold yields a maximum of 30 predictions per image.
Out of 220K image cutouts, only 39 have no radio morphology detections above this score.

\subsection{Predictions for Central Sources in Images}
\label{SEC:SelavyCentralSources}
As we acquire predictions for all 220K cutouts centred on the positions of the \textit{Selavy} catalogue components, we selectively utilize predictions from central sources in cutouts to compile the consolidated catalogue.
For each image, we initially generate a dictionary for every cutout, containing prediction details such as radio categories, bounding boxes, keypoints, and scores for all detected instances.
Subsequently, we identify predictions where bounding boxes with the highest scores overlap with the centre of each cutout. 
Remarkably, only 1.1\% of cutouts lack predicted bounding boxes at the centre, totalling approximately 2,336 cutouts without central predictions. 
A visual examination of 200 of these cutouts reveals that they all feature a faint compact radio galaxy at the centre, resulting in confidence scores below our 0.05 threshold. 
Utilizing the prediction dictionary from the remaining 98.9\% of central predictions, we proceed to update the consolidated catalogue, a process detailed in the subsequent sections.
As outlined in Table~\ref{TAB:AP1}, the AP$_{50}$ for the combined validation and test sets stands at 73.2\%, taking into account multiple predictions within an image. 
Nonetheless, upon examining these images, we observe that 99\% of central radio galaxies exhibit an IoU greater than 0.5, and 97.2\% of central radio galaxies boast an IoU exceeding 0.7. 
Additionally, in terms of keypoint detections, we observe that 98\% of central radio galaxies have a keypoint position within $<3^{\prime \prime}$ of the CatWISE host in the evaluation set, and 77\% of keypoints are $<1^{\prime \prime}$ away from the CatWISE host location.
This indicates that, for the majority of central radio galaxies, the predicted bounding boxes and keypoint positions align well with the ground truth.

\begin{figure}[!t]
\centering
\includegraphics[width=7.cm, scale=0.5]{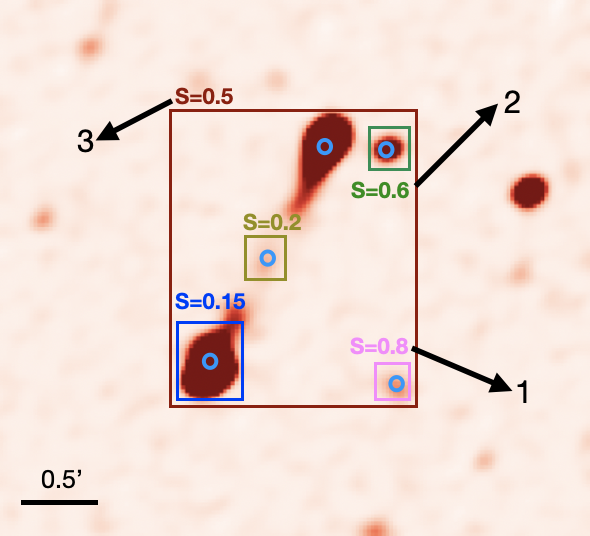}
\caption{Shown is a potential scenario in which an extended radio galaxy is accompanied by two adjacent compact radio galaxies. Blue circles represent all \textit{Selavy} catalogue components. The extended radio galaxy has a confidence score of 0.5, but the two compact radio galaxies with higher scores (0.8 and 0.6) are consolidated first, removing them from the \textit{Selavy} catalogue. Subsequently, only the three remaining components of the extended radio galaxy, encompassed by the biggest bounding box with a score of 0.5, are consolidated. As a result, the consolidated catalogue registers three radio galaxies in this particular case.} 
\label{FIG:CatPipeline2}
\end{figure}

\subsection{Cataloguing Galaxies based on Detection Scores}
\label{SEC:CataloguingWithScore}
The prediction dictionary encompasses categories, bounding boxes, keypoints, and scores for central sources. 
These galaxies are incorporated into the consolidated catalogue based on their detection scores, with higher-scoring galaxies being added first. 
To achieve this, we initiate the process by calibrating the scores for all central sources in the dictionary.
Calibration ensures that the predicted probabilities represent reliable estimates of the true probabilities.
Following the methodology outlined in \cite{gupta2021calibration}, we employ a spline-based calibration approach. 
This approach introduces a binning-free calibration measure inspired by the comparison of cumulative probability distributions in the classical Kolmogorov-Smirnov (KS) statistical test. 
Unlike traditional binning, it approximates the empirical cumulative distribution using a differentiable function achieved through splines. 
The calibration process involves finding a mapping $\gamma: [0,1] \rightarrow [0,1]$ such that $\gamma (f_k(x))$ is calibrated. 
This mapping is established through a direct correlation from the score $f_k(x)$ to $P(k|f_k(x))$ for all classes $k$. 
We refer the readers to \cite{gupta2021calibration} for more details.
The spline fitting is executed using a held-out calibration set, and the resulting calibration function is evaluated on an unseen test set.

We utilize scores and class predictions extracted from the training dataset for establishing a calibration function. The effectiveness of this calibration function is then assessed using scores and classes from the combined validation and test sets. The graphs depicting uncalibrated and calibrated scores are illustrated in Figure~\ref{FIG:ScoreCalib}.
Here, (a) presents a plot depicting the cumulative score and probability against the fractile of the test set. In (b), the same information is displayed, but the horizontal axis is distorted, resulting in a cumulative score graph that forms a straight line. This is visualized through scatter plots of cumulative (score, score) in blue and (score, probability) in orange. In the case of a perfectly calibrated network, the probability line will align as a straight line coinciding with the (score, score) line.
In (c) and (d), plots of non-cumulative scores and probabilities are presented against the fractile or score. 
The upper panel illustrates that the network is substantially overestimating the probability of the detection based on the score.
The bottom panel displays the outcomes of the calibration, revealing a notable improvement compared to the upper panel. This is evident from the lower KS error of 4.2\%, in contrast to 7.3\%. 
It is also important to highlight that the enhanced calibration is achieved without compromising accuracy (or probability).

We employ the calibration function on the scores of central sources within the prediction dictionary. 
Subsequently, we arrange the scores and their associated category, bounding box, and keypoint predictions in descending order. 
The bounding box predictions are then utilized in a descending order of scores to group all \textit{Selavy} catalogue components within each box and collectively add them to the consolidated catalogue. 
Simultaneously, we remove the corresponding rows of these components from the \textit{Selavy} catalogue. 
Since boxes with higher scores are prioritized, the components already removed from the \textit{Selavy} catalogue do not duplicate in the consolidated catalogue. 
This process is iterated for all central source predictions in the dictionary. 
Ultimately, when no components remain in the \textit{Selavy} catalogue, the cataloguing process concludes.

\subsection{Predictions for Large extended Radio Galaxies}
\label{SEC:MultipleInBigBox}
In a few cases, there are compact radio galaxies in the proximity of a large extended radio galaxy.
The Gal-DINO model would generally predict a large bounding box encompassing all the components of the extended radio galaxy, but would also predict bounding boxes for the neighbouring radio galaxies.
In such a scenario, the cataloguing process explained in the previous sections will tend to automatically add components with higher confidence scores to the consolidated catalogue but as it also removes those from the \textit{Selavy} catalogue, this leads to separating the neighbouring radio galaxies from the large extended radio galaxy.
As an example, Figure~\ref{FIG:CatPipeline2} shows such a possible scenario, where an extended radio galaxy has two compact radio galaxies in its proximity.
All \textit{Selavy} catalogue components in the figure are represented by blue circles.
The confidence score for the extended radio galaxy is 0.5 (with the biggest bounding box), however, as the two compact radio galaxies have higher scores, the galaxy with a score of 0.8 is consolidated first and removed from the \textit{Selavy} catalogue.
This is followed by the second compact radio galaxy with a score of 0.6.
As both compact radio galaxies are consolidated and removed from \textit{Selavy} catalogue, only the three remaining components are consolidated for the extended radio galaxy with the biggest box with a score of 0.5 encompassing them.
Thus the consolidated catalogue records three radio galaxies in this scenario.

Note that the predictions with scores 0.2 and 0.15 do not matter here as these components are already consolidated and removed from the \textit{Selavy} catalogue due to their presence in the bigger box with a higher score of 0.5.
A visual inspection of several such extended radio galaxies shows that this process works for most of the extended radio galaxies with neighbouring compact radio galaxies.
However, for a handful of extended radio galaxies, the neighbouring compact radio galaxy gets consolidated due to its lower score as compared to the bigger box's score that encloses both extended and compact radio galaxies.
Future work with the EMU main survey should revisit this issue using a larger sample size, which can be addressed separately with a different computer vision strategy.
Additionally, addressing the issue posed by very large extended radio galaxies with extensions surpassing $8^\prime$ should be a focus of future work with the EMU main survey.
These galaxies, characterized by their limited representation in the present work due to their scarcity in the EMU-PS, were not incorporated into the training of the Gal-DINO model.

\begin{figure}[!t]
\centering
\vspace{0.2cm}
\includegraphics[width=6.5cm, scale=0.5]{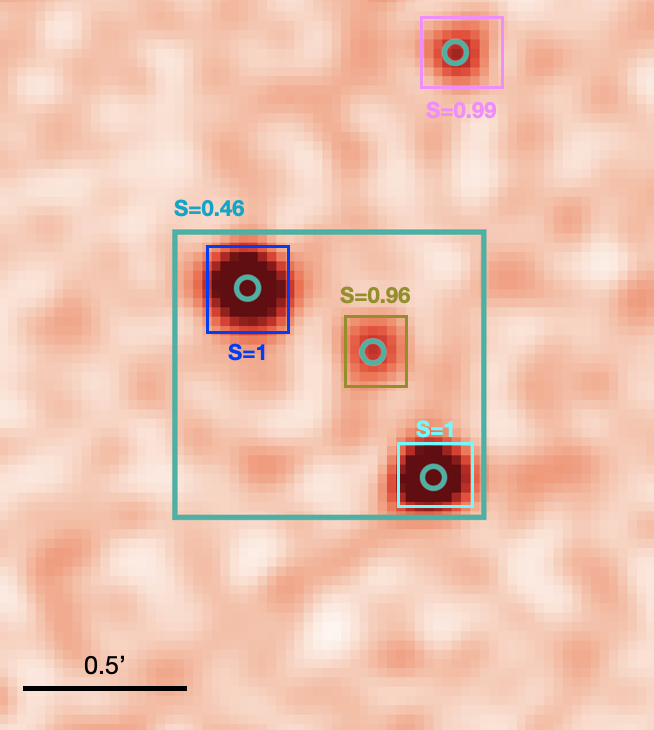}
\caption{An example where the larger box prediction, scoring 0.48, is deemed redundant. The existence of smaller boxes within implies that these contain three compact radio galaxies, each with higher scores, rather than a single, bent FR-II radio galaxy within the larger bounding box.} 
\label{FIG:CatPipeline3}
\end{figure}

\begin{figure}[!t]
\centering
\includegraphics[width=7.5cm, scale=0.5]{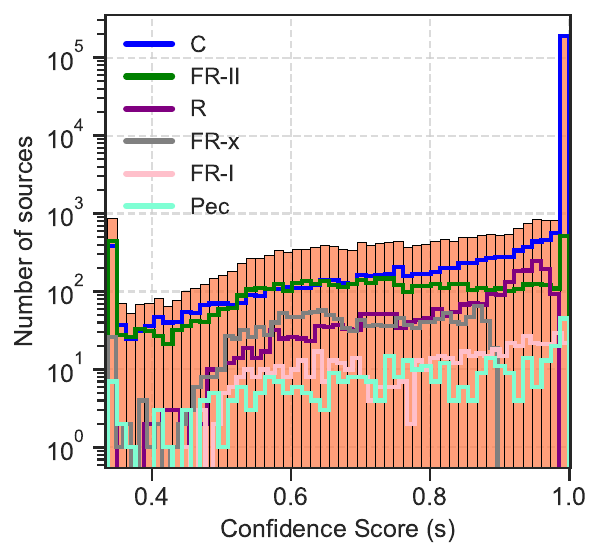}
\includegraphics[width=7.2cm, scale=0.5]{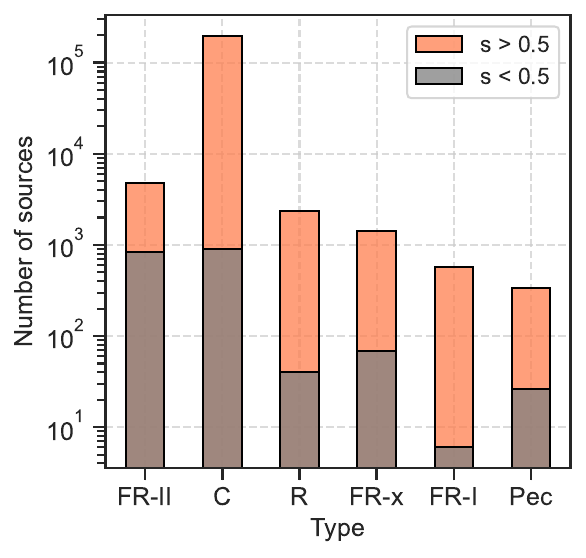}
\caption{The top panel depicts the distributions of all radio galaxies (solid bars) and subsets categorized by classification types (coloured contours) based on prediction confidence scores.
The bottom panel displays the distribution across different types as predicted by the Gal-DINO model above and below scores of 0.5. Approximately 99.1\% of the radio galaxies in our consolidated catalogue have a score larger than 0.5. Notably, among the galaxy types, FR-II exhibits the largest fraction of galaxies with scores below 0.5; for further insights, refer to Section~\ref{SEC:ConsolidatedCat}.} 
\label{FIG:ScoreHists}
\end{figure}

\section{Catalogue Descprition}
\label{SEC:CatDescription}
The final catalogue comprises 211,625 radio galaxies, and 73\% of these have a counterpart in the CatWISE infrared catalogue within $3^{\prime \prime}$. 
Additionally, cross-matched counterparts from the optical surveys have been incorporated into the catalogue.
Table~\ref{TAB:column_description1} provides detailed descriptions of the catalogue columns.
This section elaborates on the specifics of the catalogue.

\begin{figure*}[!t]
\centering
\vspace{0.2cm}
\includegraphics[width=8.5cm, scale=0.5]{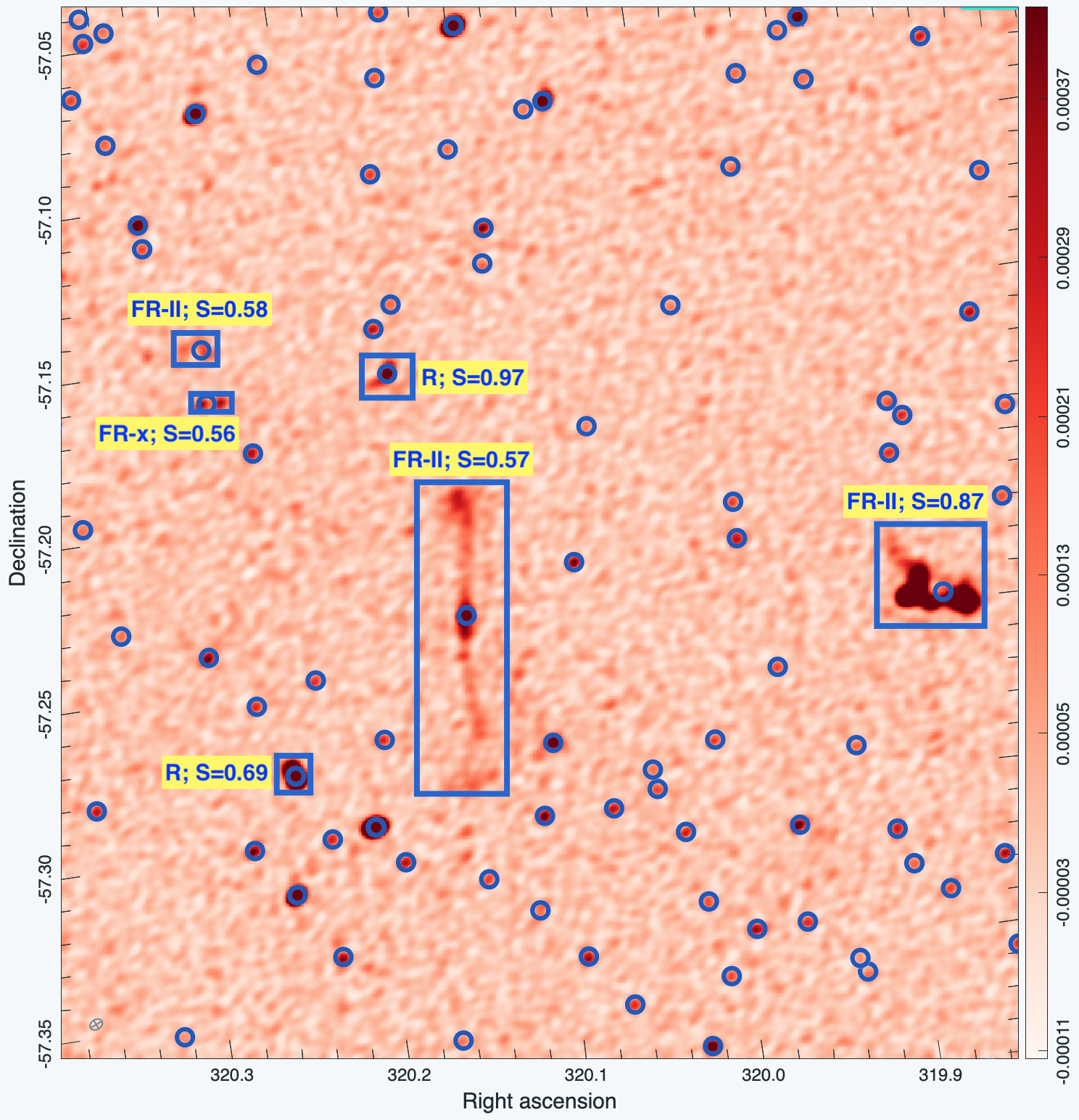}
\includegraphics[width=8.5cm, scale=0.5]{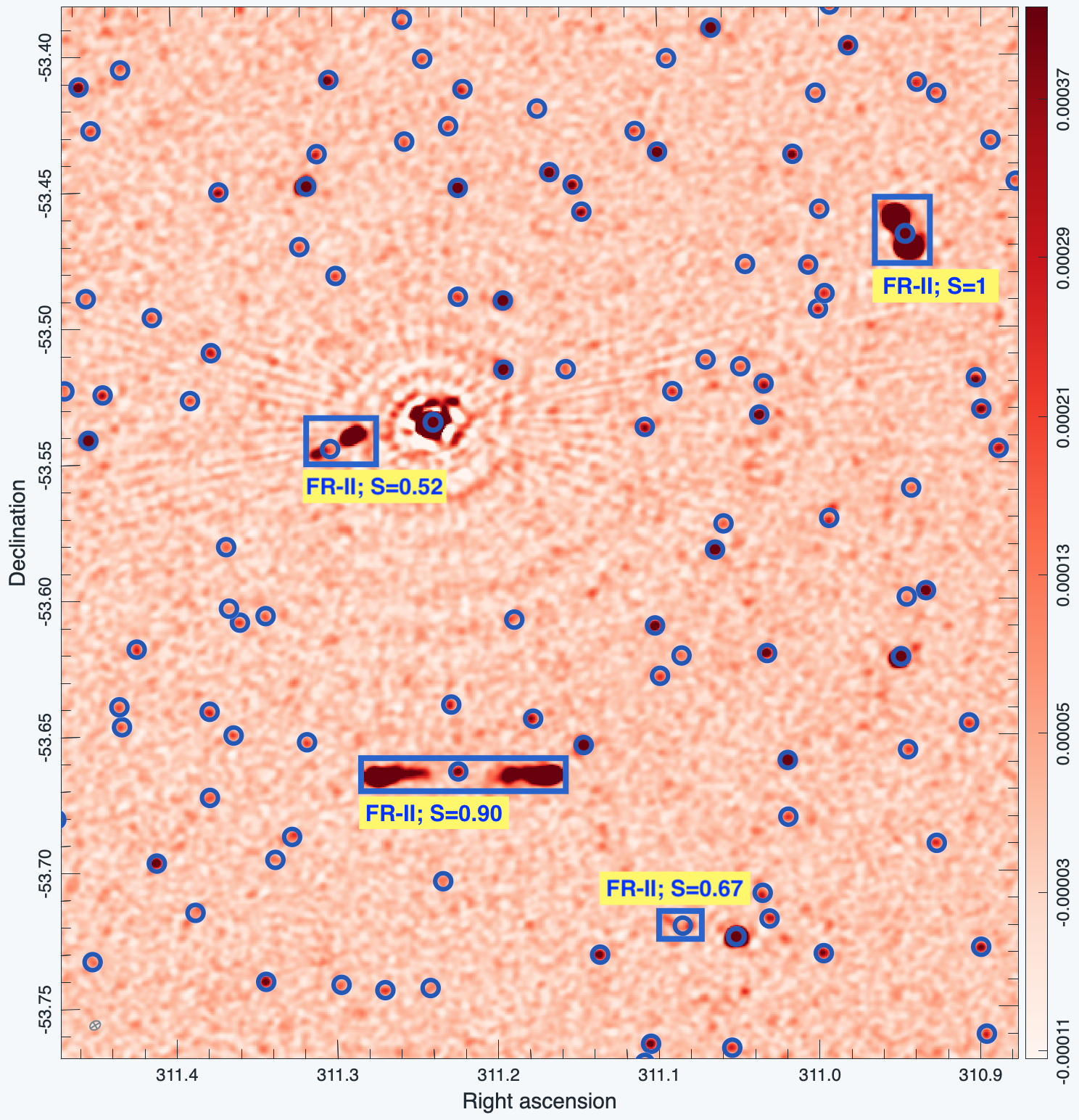}
\includegraphics[width=8.5cm, scale=0.5]{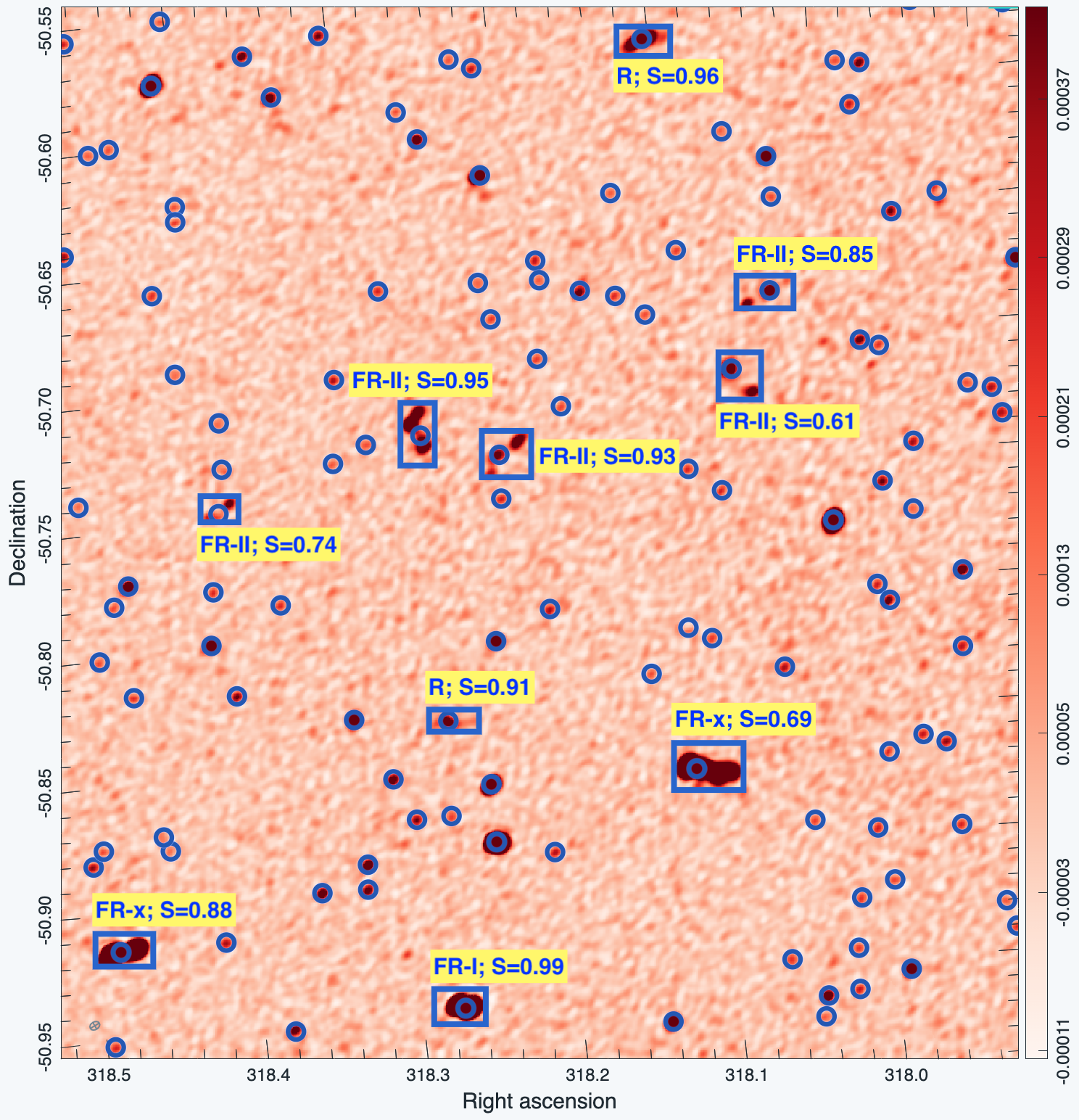}
\includegraphics[width=8.5cm, scale=0.5]{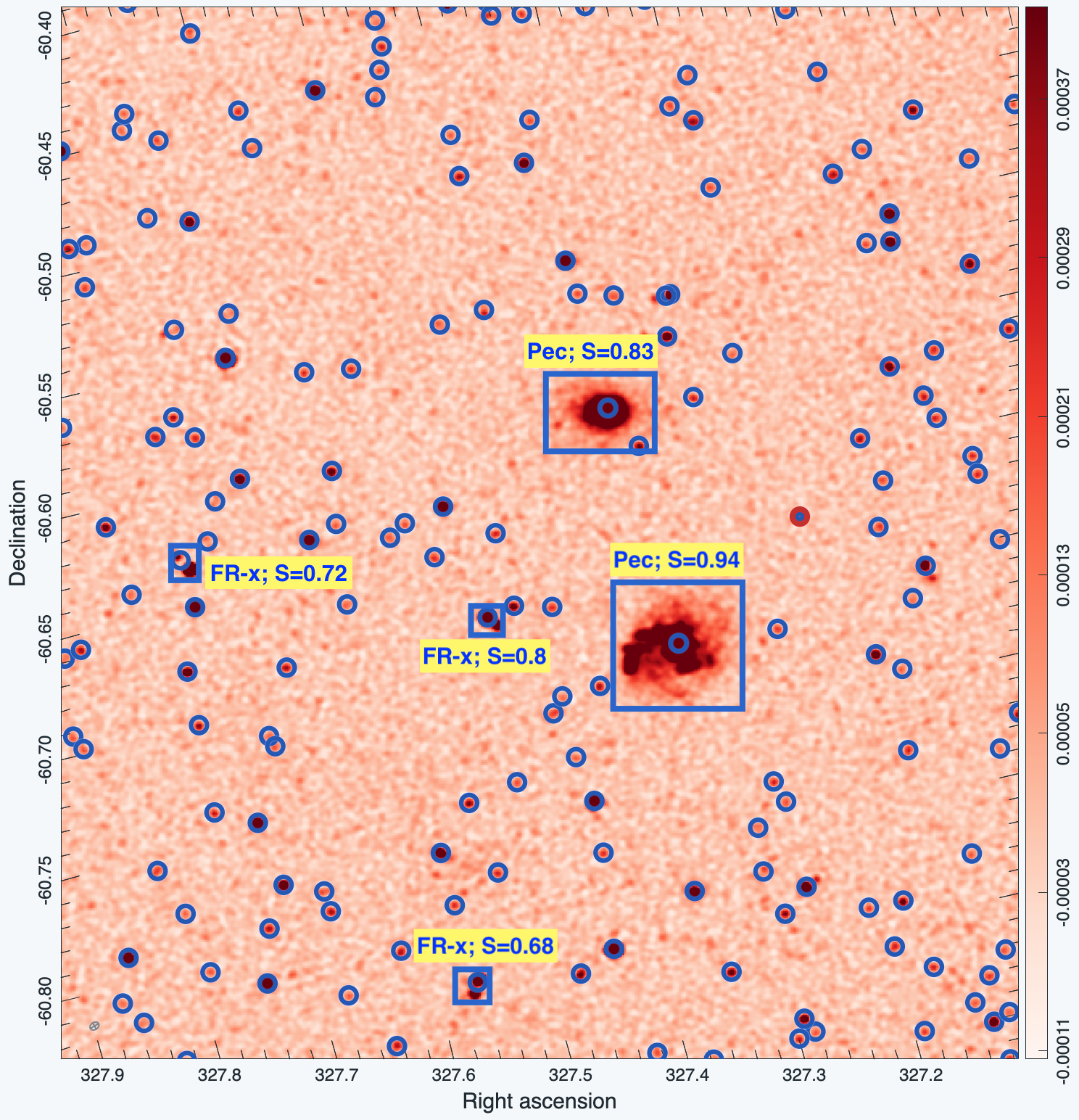}
\caption{Examples of the catalogue galaxies overlaid on radio images, showcasing their host positions, bounding boxes, and classifications derived from both the Gal-DINO and catalogue construction pipelines. Blue rectangles and accompanying text denote the bounding boxes and classification types with confidence scores for the extended radio galaxies. The positions of these extended radio galaxies are marked by blue circles. For brevity, we omit the presentation of bounding boxes for compact radio galaxies, which are solely indicated by blue circles.} 
\label{FIG:CatImageExamples}
\end{figure*}

\subsection{Consolidated Catalogue}
\label{SEC:ConsolidatedCat}
The upper panel in Figure~\ref{FIG:ScoreHists} displays a histogram depicting the prediction confidence scores for all radio galaxies.
Out of the 211,625 radio galaxies in the catalogue, 91.1\% of them exhibit a confidence score of 1, and 99.1\% are identified with a score greater than 0.5. 
As detailed in Section~\ref{SEC:SelavyCentralSources}, there are 2,336 cutouts without central source predictions surpassing the confidence score of 0.05. 
Among these, 262 are incorporated through consolidation.
As all the remaining 2,074 correspond to single-component islands in the \textit{Selavy} catalogue as well, we include them as compact radio galaxies at the end of the consolidated catalogue.
The catalogue comprises 201,211 compact radio galaxies and 10,414 extended radio galaxies.  
Among the extended radio galaxies, there are 582 FR-I, 5,602 FR-II, 1,494 FR-X, and 2,375 R radio galaxies as well as 361 Pec morphologies.
Among the 0.9\% (1,877) of galaxies with confidence scores below 0.5, the majority are associated with C (898) and FR-II (838) galaxies, as depicted in the lower panel of Figure~\ref{FIG:ScoreHists}. This corresponds to only 0.05\% of the total compact radio galaxies and 15\% of the total FR-II galaxy predictions below a score of 0.5. 
Future investigations should focus on a more detailed examination of these FR-II galaxies with scores below 0.5. 
In our visual examinations of 100 systems classified as FR-II with scores below 0.5, we noted that, in several cases, their extensive bounding boxes encapsulate multiple compact radio galaxies, each with a higher score than the score assigned to the larger bounding box. 
As described in Section~\ref{SEC:MultipleInBigBox} and illustrated in Figure~\ref{FIG:CatPipeline2}, galaxies with higher scores are prioritized for consolidation in the catalogue.
We observe that several bounding boxes classified as FR-II encompass adjacent compact radio galaxies rather than genuine FR-II galaxies; however, 30 out of 100 are indeed real FR-II galaxies.
In Figure~\ref{FIG:CatPipeline3}, an instance of this situation is demonstrated, where the larger box prediction, with a score of 0.48, proves to be redundant. 
The presence of smaller boxes within suggests that these encompass three compact radio galaxies, each with higher scores, rather than indicating a single FR-II galaxy.
Also after examining 100 randomly chosen predictions corresponding to 85\% of FR-II classifications with scores above 0.5, we found that 90 of them are indeed authentic FR-II galaxies. 
However, 9 are FR-I or FR-X and one bounding box encompasses multiple compact radio sources.
This indicates that the classification of FR-II galaxies can be relied upon for specific tasks, with more reliable classifications having scores above 0.5.
It is crucial to emphasize that this does not impact the overall catalogue, as bounding boxes for low-scoring FR-II galaxies often have high-scoring compact radio galaxies, and these are already accounted for in the consolidated catalogue (as explained in Section~\ref{SEC:MultipleInBigBox}).
The lower panel of Figure~\ref{FIG:ScoreHists} illustrates the distribution of galaxies across the six prediction class types.
Following this, we identify 200,313 compact radio galaxies, 576 FR-I galaxies, 4,764 FR-II galaxies, 1,425 FR-x galaxies, 2,335 R galaxies, and 335 Pec morphologies with model confidence scores above 0.5.
Note that the accuracy of predictions relies on the appropriate selection of the confidence score, which should align with the specific requirements of the scientific case.

\begin{table}[!ht]
\centering
\resizebox{\textwidth}{!}{%
\begin{tabular}{ll}
\hline
\textbf{Column} & \textbf{Description} \\
\hline
EMU\_PS\_source\_name & Name of the EMU-PS radio galaxy \\
EMU\_PS\_ra\_deg & Right Ascension of the EMU-PS galaxy (degrees) \\
EMU\_PS\_dec\_deg & Declination of the EMU-PS galaxy (degrees) \\
EMU\_PS\_flux\_peak & Peak flux density of the EMU-PS radio galaxy (mJy/beam)\\
EMU\_PS\_flux\_int & Integrated flux density of the EMU-PS radio galaxy (mJy)\\
EMU\_PS\_img\_id & Image identifier of the EMU-PS galaxy \\
EMU\_PS\_filename & Filename associated with the EMU-PS galaxy \\
EMU\_PS\_radio\_bbox & Bounding box of the EMU-PS radio galaxy (pixels) \\
EMU\_PS\_ra\_cen\_bbox & Right Ascension centre of the EMU-PS bounding box (degrees)\\
EMU\_PS\_dec\_cen\_bbox & Declination centre of the EMU-PS bounding box (degrees)\\
EMU\_PS\_type & Type classification of the EMU-PS galaxy \\
EMU\_PS\_score & Classification score of the EMU-PS galaxy \\
EMU\_PS\_n\_selavy\_components & Number of \textit{Selavy} components for the galaxy \\
selavy\_island\_id & Identifier for \textit{Selavy} island \\
selavy\_component\_id & Identifier for \textit{Selavy} component \\
selavy\_component\_name & Name of \textit{Selavy} component \\
selavy\_ra\_hms\_cont & Right Ascension of \textit{Selavy} component (HMS)\\
selavy\_dec\_hms\_cont & Declination of \textit{Selavy} component (DMS)\\
selavy\_ra\_deg\_cont & Right Ascension of \textit{Selavy} component (degrees)\\
selavy\_dec\_deg\_cont & Declination of \textit{Selavy} component (degrees)\\
selavy\_ra\_err & Right Ascension error of \textit{Selavy} component (arcsec)\\
selavy\_dec\_err & Declination error of \textit{Selavy} component (arcsec)\\
selavy\_flux\_peak & Peak flux density of \textit{Selavy} component (mJy/beam)\\
selavy\_flux\_peak\_err & Error in peak flux density of \textit{Selavy} component (mJy/beam)\\
selavy\_flux\_int & Integrated flux density of \textit{Selavy} component (mJy)\\
selavy\_flux\_int\_err & Error in integrated flux density of \textit{Selavy} component (mJy)\\
selavy\_spectral\_index & Spectral index of \textit{Selavy} component \\
catwise\_source\_id & Identifier for CatWISE source \\
catwise\_source\_name & Name of CatWISE source \\
catwise\_ra & Right Ascension of CatWISE source (degrees)\\
catwise\_dec & Declination of CatWISE source (degrees)\\
w1mag & CatWISE W1 magnitude \\
w1sigm & CatWISE W1 magnitude error \\
w2mag & CatWISE W2 magnitude \\
w2sigm & CatWISE W2 magnitude error \\
catwise\_separation & Separation between the radio galaxy and CatWISE source (arcsecond)\\
des\_coadd\_object\_id & DES coadd object identifier \\
des\_ra & Right Ascension of DES source (degrees)\\
des\_dec & Declination of DES source (degrees)\\
des\_mag\_auto\_g & DES magnitude in the g-band \\
des\_magerr\_auto\_g & Error in DES magnitude in the g-band \\
des\_mag\_auto\_r & DES magnitude in the r-band \\
des\_magerr\_auto\_r & Error in DES magnitude in the r-band \\
des\_mag\_auto\_i & DES magnitude in the i-band \\
des\_magerr\_auto\_i & Error in DES magnitude in the i-band \\
des\_mag\_auto\_z & DES magnitude in the z-band \\
des\_magerr\_auto\_z & Error in DES magnitude in the z-band \\
des\_mag\_auto\_y & DES magnitude in the y-band \\
des\_magerr\_auto\_y & Error in DES magnitude in the y-band \\
des\_separation & Separation between the CatWISE source and DES source (arcsecond)\\
scosID & SuperCosmos identifier \\
scos\_htmID & SuperCosmos HTM identifier \\
scos\_ra & Right Ascension of SuperCosmos source (degrees)\\
scos\_dec & Declination of SuperCosmos source (degrees)\\
scos\_zPhoto\_ANN & SuperCosmos redshift from ANN \\
scos\_zPhoto\_Corr & Corrected SuperCosmos redshift \\
scos\_separation & Separation between the CatWISE source and SuperCosmos source (arcsecond)\\
desi\_ID & DESI identifier \\
desi\_RA & Right Ascension of DESI source (degrees)\\
desi\_DEC & Declination of DESI source (degrees)\\
desi\_TYPE & Type classification of DESI source \\
desi\_photo\_z & Photometric redshift of DESI source \\
desi\_photo\_zerr & Error in photometric redshift of DESI source \\
desi\_spec\_z & Spectroscopic redshift of DESI source \\
desi\_separation & Separation between the CatWISE source and DESI (arcsecond)\\
\hline
\end{tabular}%
}
\caption{Description of columns in the catalogue (best viewed in a PDF reader).}
\label{TAB:column_description1}
\end{table}

The consolidated catalogue includes the following columns: EMU\_PS\_source\_name (e.g., EMU PS J203428.9-583922\footnote{Nomenclature criterion details: \url{https://cdsarc.cds.unistra.fr/viz-bin/Dic?/4848739}} where `P' stands for pilot survey, and `S' stands for source), derived from the Right Ascension (RA) and Declination (Dec) of the consolidated radio galaxies. 
The RA (EMU\_PS\_ra\_deg) and Dec (EMU\_PS\_dec\_deg) represent the detected keypoint positions. 
It is important to note that we use keypoint positions for RA and Dec, rather than the position of the peak radio emission inside the predicted bounding box encompassing extended radio galaxies. 
This choice is made to avoid using the positions of radio lobes for RA and Dec, as they can have higher flux compared to the central radio source.
As keypoint detection is performed on cutouts with 7 radio and 1 infrared channels, as done for locating bounding boxes. 
Consequently, the keypoints denote positions where distinctive features in both radio and infrared images contribute to their predictive capabilities. 
While keypoints in the training data correspond to known CatWISE sources, typically situated near the central core or the ridge of collimated radio emission connecting the lobes, the model predictions rely on features extracted from both radio and infrared images. 
In scenarios where no infrared emission is present near the radio components, the model tends to predict a keypoint near the central radio peak or the ridge, rather than artificially placing it randomly in the infrared image.
Therefore, we utilize keypoint positions for RA and Dec instead of radio peak positions for extended radio galaxies. 
Despite there being only one radio peak for compact radio galaxies, for consistency, we use keypoint positions for both extended and compact radio galaxies.

The flux density columns, EMU\_PS\_flux\_peak and EMU\_PS\_flux\_int, are sourced from the Selavy catalogue, where a signal-to-noise ratio of $\geq5$ is applied to select radio sources.
The peak flux density is obtained from the maximum flux density of consolidated components, while the integrated flux density is the sum of these components.
EMU\_PS\_img\_id and EMU\_PS\_filename serve as identifiers and names for the cutouts from which the galaxy was consolidated into our catalogue. 
These identifiers correspond to component IDs in the \textit{Selavy} catalogue, facilitating the viewing of consolidated galaxy cutouts.
EMU\_PS\_radio\_bbox provides the dimensions of the detected bounding box for the radio galaxy in pixels, including information about the minimum x and y coordinates, height, and width. 
The height and width, measured in pixels and parallel to RA and Dec, can be used to determine the total extent of the radio galaxies, with each pixel equivalent to 2$^{\prime \prime}$.
Above an integrated flux of 10 mJy for extended radio galaxies, we measure the largest angular size (LAS) using the height and width, yielding a median LAS of $93.6^{\prime \prime}$ with a standard deviation of $50.4^{\prime \prime}$. 
This is consistent with \cite{williams19}, where employing the same flux threshold at 944 MHz (assuming a spectral index of -0.7) and selecting radio sources featuring multiple components, we find a median LAS of $99.6^{\prime \prime}$ with a standard deviation of $74.9^{\prime \prime}$. 
EMU\_PS\_ra\_cen\_bbox and EMU\_PS\_dec\_cen\_bbox denote the box centroids in RA and Dec, serving as box positions for specific tests.
EMU\_PS\_type and EMU\_PS\_score indicate the predicted classification types and confidence scores, respectively.
Figure~\ref{FIG:CatImageExamples} illustrates examples from the consolidated catalogue overlaid on EMU-PS radio images. 
It showcases RA and Dec positions with blue circles, extended radio galaxy bounding boxes with blue rectangles, and classification types with confidence scores for different radio galaxies exhibiting various extents.
The last column, EMU\_PS\_n\_selavy\_components, provides information about the number of \textit{Selavy} components for each consolidated radio galaxy.
Table~\ref{TAB:Catalog} displays the initial 60 rows of the catalogue, showcasing the EMU-PS fields.

\begin{figure}[!t]
\centering
\includegraphics[width=7.5cm]{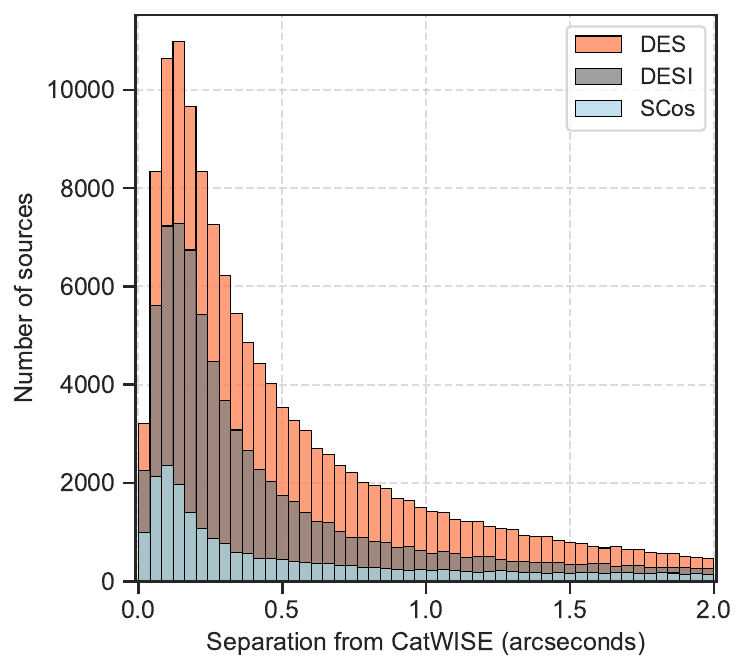}
\includegraphics[width=7.1cm, scale=0.5]{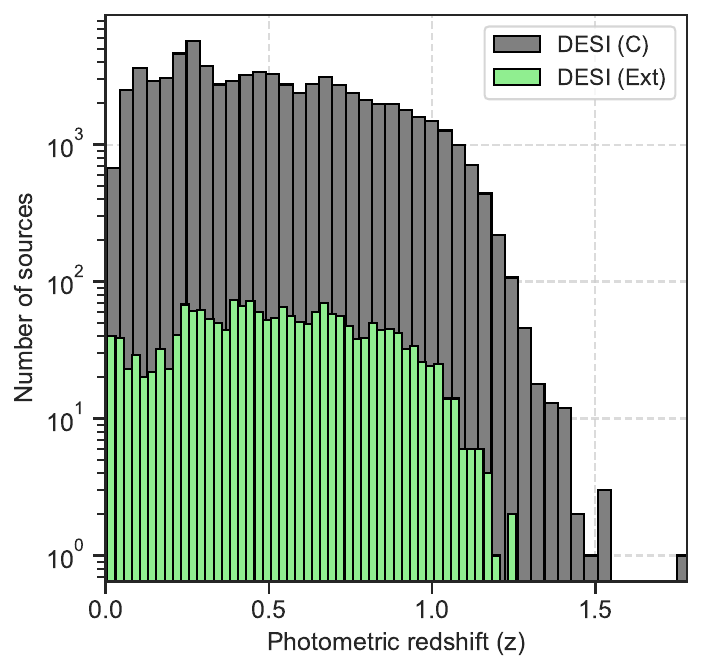}
\caption{Displayed are the numbers of DES, DESI, and SCos counterparts within 2$^{\prime \prime}$ of the CatWISE position (top panel). 
The photometric redshift distributions are based on the DESI legacy surveys for compact and extended radio galaxy counterparts (bottom panel). 
Both these plots include galaxies with a Gal-DINO confidence score greater than 0.5.} 
\label{FIG:CounterpartStats}
\end{figure}

\subsection{Comparison to $Selavy$}
\label{SEC:selavyCompare}
The compilation of the consolidated catalogue relies on the sources identified by \textit{Selavy}.
As mentioned in Section~\ref{SEC:ASKAP}, the purpose of the \textit{Selavy} algorithm is not to generate radio galaxy catalogues.
In the \textit{Selavy} process, pixels above a specific brightness threshold are initially detected and grouped into islands. 
Following this, components are fitted to each island, ultimately forming the component catalogue.
Out of 198,216 \textit{Selavy} islands, 178,921 are single-component, and the remaining 19,295 are multi-component with a total of 41,181 components. 
We record the selavy\_island\_id and selavy\_component\_id for each galaxy in the consolidated catalogue. 
As detailed in Section~\ref{SEC:ConsolidatedCat}, the consolidated catalogue comprises 211,625 radio galaxies, representing an increase of 13,409 radio galaxies compared to the 198,216 islands in \textit{Selavy}.
Among these, the selavy\_island\_id in our consolidated catalogue consists of 204,533 single-component and 7,092 multi-component \textit{Selavy} island IDs. 
This indicates the presence of 204,533 distinct islands, compared to the 178,921 islands with a single component in the consolidated catalogue. 
In essence, \textit{Selavy} seems to underestimate the number of single-component islands by 12.5\%.
Regarding the 7,092 multi-component island IDs, 6,901 share the same set of 15,161 components as identified by \textit{Selavy}, indicating correct associations for these islands. 
The remaining 191 multi-component island IDs have 407 components all from different \textit{Selavy} islands. 
Overall, \textit{Selavy} correctly associates multiple components for 6,901 out of 19,295 islands, with a correct association rate of 35.7\% for the multi-component islands.

Furthermore, we incorporate additional attributes about the \textit{Selavy} catalogue. 
Each \textit{Selavy} component is uniquely identified by its assigned name, denoted as selavy\_component\_name. 
The celestial coordinates, namely Right Ascension (RA) and Declination (Dec), are encapsulated in the fields selavy\_ra\_deg\_cont and selavy\_dec\_deg\_cont, respectively. 
Information regarding the peak and integrated flux density of the radio signal emitted by these \textit{Selavy} components can be found in selavy\_flux\_peak and selavy\_flux\_int. 
Additionally, the spectral index, characterizing the frequency dependence of the radio emission, is detailed in selavy\_spectral\_index (see Figure 11 in \cite{norris21} also).

\begin{table}[!t]
\centering
\resizebox{\textwidth}{!}{%
\begin{tabular}{lrrr}
\hline
Criterion    & \# All (\%) & \# Compact (\%) & \# Extended (\%) \\
\hline
EMU PS (score>0.33)                    & 211,625 (100\%)  & 201,211 (100\%)  & 10,414 (100\%) \\
EMU PS-CatWISE (<3$^{\prime \prime}$)     & 154,320 (73\%)   & 149,789 (74\%)   & 4,531  (44\%) \\
CatWISE-DES (<2$^{\prime \prime}$)         & 135,266 (64\%)   & 131,464 (65\%)   & 3,802  (37\%) \\
CatWISE-DESI (<2$^{\prime \prime}$)        & 75,904  (36\%)   & 73,812  (37\%)   & 2,092  (20\%) \\
CatWISE-SCos (<2$^{\prime \prime}$)        & 22,242  (11\%)   & 21,822  (11\%)   & 420    (4\%) \\
\hline
EMU PS (score>0.5)                     & 207,674 (100\%)  & 198,239 (100\%)  & 9,435 (100\%) \\
EMU PS-CatWISE (<3$^{\prime \prime}$)     & 152,775 (74\%)   & 148,593 (75\%)   & 4,182 (44\%) \\
CatWISE-DES (<2$^{\prime \prime}$)         & 133,964 (65\%)   & 130,445 (66\%)   & 3,519 (37\%) \\
CatWISE-DESI (<2$^{\prime \prime}$)        & 75,216  (36\%)   & 73,268  (37\%)   & 1,948 (21\%) \\
CatWISE-SCos (<2$^{\prime \prime}$)        & 22,122  (11\%)   & 21,732  (11\%)   & 390   (4\%) \\
\hline
\end{tabular}%
}
\caption{Number of CatWISE, DES, DESI and SCos counterparts for radio galaxies. The cross-matching process involves CatWISE sources matched with radio galaxies within a 3$^{\prime \prime}$ search radius while cross-matching with DES, DESI, and SCos catalogues utilizes these CatWISE sources within a 2$^{\prime \prime}$ radius.
The upper and lower section of the table presents the number of galaxies and the percentage of cross-matches with confidence scores exceeding 0.33 and 0.5, respectively. 
The numbers and percentages refer to the sources remaining after applying all of the previous criteria.
The columns display these statistics for all, compact, and extended radio galaxies.}
\label{TAB:MultiwavelengthNumbers}
\end{table}

\subsection{Multiwavelength Counterparts}
\label{SEC:MultiwavelengthCounterparts}
We cross-identify radio galaxies in the consolidated catalogue with the infrared and optical catalogues. 
To do this, we first cross-match the EMU\_PS\_ra\_deg and EMU\_PS\_dec\_deg coordinates with the CatWISE catalogue. 
As explained in Section~\ref{SEC:ConsolidatedCat}, these coordinates originate from the keypoint positions, and we do not utilize radio peak flux or bounding box centroids for this cross-identification.
We use a 3$^{\prime \prime}$ search radius around radio galaxies to identify counterparts. 
At this 3$^{\prime \prime}$ search radius, the false identification rate is expected to be around 8\% \citep[see Figure 12 in][]{norris21}. 
Table~\ref{TAB:MultiwavelengthNumbers} displays the number of radio galaxies within this search radius. 
We find that 73\% of radio galaxies (including the 8\% with possible false identifications) have counterparts in the CatWISE catalogue. 
Consequently, the consolidated catalogue lacks multiwavelength information for the remaining 27\% of radio sources.
Among the radio galaxies classified by the Gal-DINO model as compact and all extended, we find cross-matches for 74\% and 44\%, respectively. 
Future work should investigate whether a larger search radius could be more optimal for identifying counterparts for extended radio galaxies, aiming to increase completeness while keeping the false identification rate low. 
In another scenario, when looking only for counterparts of galaxies with a GalDINO confidence score greater than 0.5, we find similar percentages of cross-matches when applying a minimum score threshold of 0.33.
Note that we have chosen this straightforward approach to identify radio galaxy counterparts, which yields acceptable false identification rates. 
Subsequent research should explore the cross-matching of radio galaxies by leveraging photometric properties, such as utilizing stellar mass estimates for the matching process \citep[see, e.g.,][]{gupta20b}, employing other likelihood ratio-based methods \citep[e.g.,][]{mcalpine12, weston18, williams19, gordon23}, filtering out contaminating stars using WISE colours or proper motion data and understanding the contamination in identification caused by gravitational lensing.

\begin{figure}[!t]
\centering
\includegraphics[width=6.7cm]{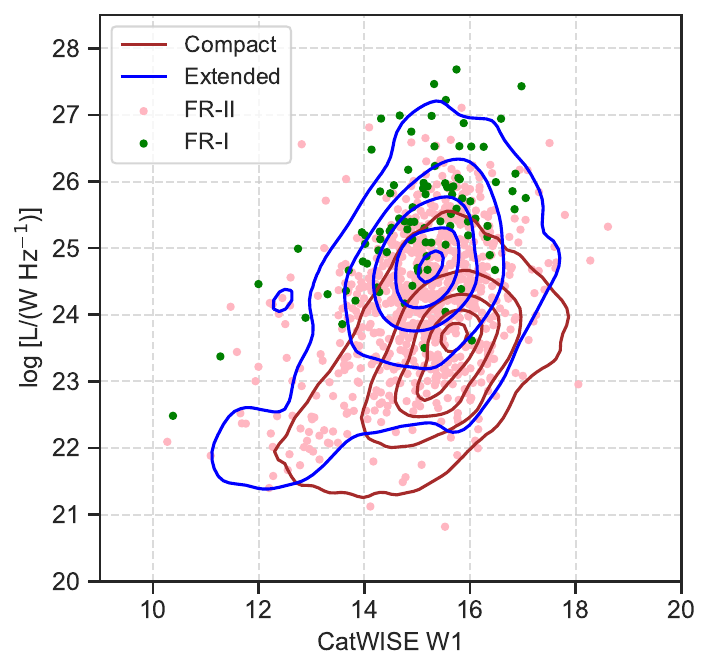}
\includegraphics[width=6.9cm, scale=0.5]{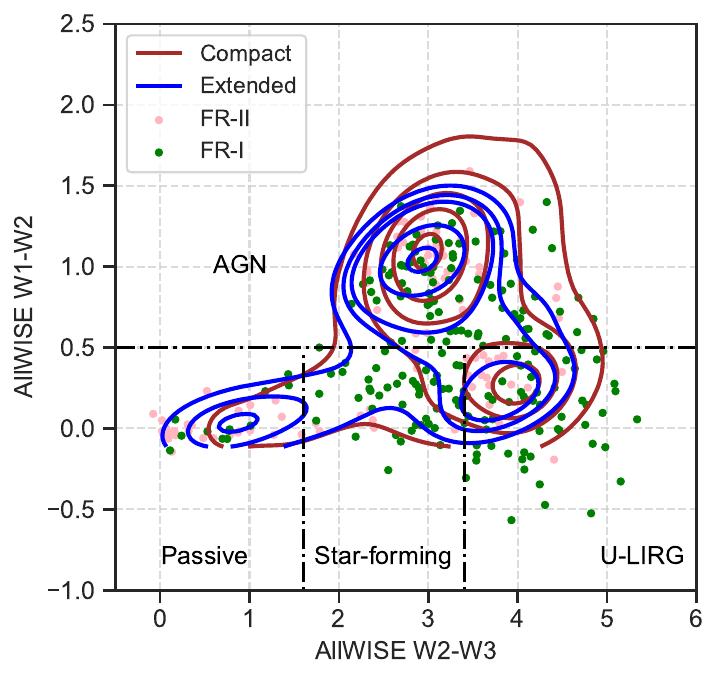}
\includegraphics[width=6.9cm, scale=0.5]{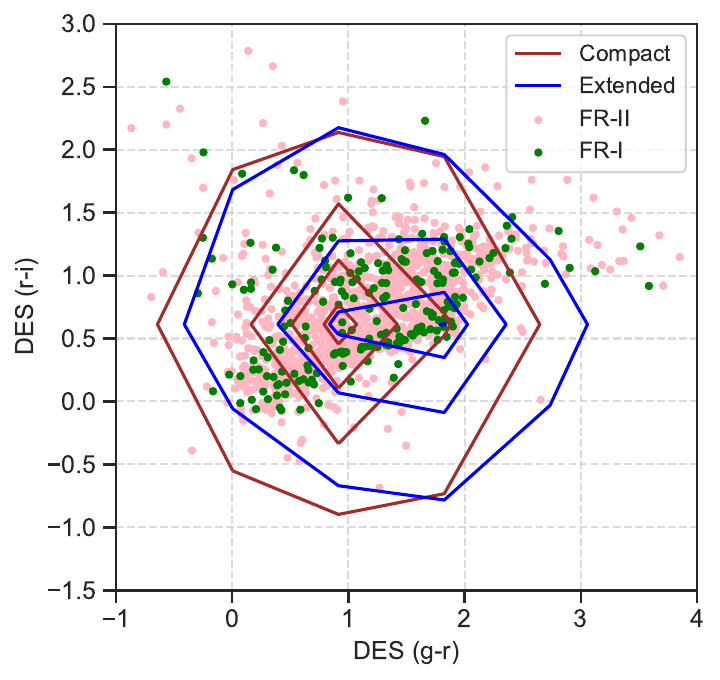}
\caption{The distributions of multiwavelength counterparts for compact and extended radio galaxies, as well as FR-I and FR-II type radio galaxies, are shown in three panels. The top panel displays contours for radio luminosity and CatWISE magnitudes, the middle panel illustrates the infrared colour-colour plot for AllWISE counterparts, emphasizing the dominance of AGN above an integrated radio flux of 5 mJy. The bottom panel exhibits the colour-colour plot for DES radio galaxy counterparts. All plots include galaxies with a Gal-DINO confidence score greater than 0.5, and contours describe 5, 25, 50, 75, and 95 percentile levels. Further details and discussion can be found in Section~\ref{SEC:MultiwavelengthCounterparts}.} 
\label{FIG:CounterpartStats2}
\end{figure}

Next, we cross-match CatWISE sources that are within a 3$^{\prime \prime}$ radius of the radio galaxies with the optical catalogues. 
We use photometric catalogues from the Dark Energy Survey Data Release 2 \citep[DES DR2][]{abbott21}, the Dark Energy Spectroscopic Instrument Legacy Survey catalogue from Data Release 8 \citep[DESI DR8][]{zou19}, and the WISE x SuperCOSMOS photometric catalogue \citep[SCos][]{bilicki16} to cross-match CatWISE sources. 
Following \citep[Figure 13 in][]{norris21}, we search for counterparts in these optical catalogues within a 2$^{\prime \prime}$ radius of CatWISE, where the false identification rate is approximately 7\%. 
Table~\ref{TAB:MultiwavelengthNumbers} and the top panel of Figure~\ref{FIG:CounterpartStats} show the number of CatWISE counterparts in DES, DESI, and SCos. 
The median separation distance for these galaxies with the CatWISE sources is found to be 0.34$^{\prime \prime}$, 0.27$^{\prime \prime}$, and 0.3$^{\prime \prime}$ for DES, DESI, and SCos catalogues. 
We find that 65\%, 36\%, and 11\% of CatWISE sources within a 3$^{\prime \prime}$ radius of radio galaxies have counterparts in DES, DESI, and SCos catalogues, respectively. 
The same percentage of counterparts is found when a confidence score limit of 0.5 is applied. 
The bottom panel of Figure~\ref{FIG:CounterpartStats} shows the number of compact (grey) and extended radio galaxy (green) counterparts with a score larger than 0.5 as a function of photometric redshift from the DESI legacy survey. 
The median redshift ($z$) for compact and extended radio galaxy counterparts is 0.46 and 0.53, respectively, with a maximum redshift of $1.79\pm0.28$ for compact and $1.26\pm0.22$ for extended. 
We find 4,525 compacts and 82 extended radio galaxy counterparts above a redshift of 1. 

The three panels in Figure~\ref{FIG:CounterpartStats2} depict the relationships between compact (brown contours) and extended radio galaxies (blue contours) and their counterparts in infrared and optical, with Gal-DINO confidence scores larger than 0.5.
In the top panel, radio luminosity contours are displayed at 5, 25, 50, 75, and 95 percentile levels relative to the CatWISE W1 magnitude. Extended radio galaxies exhibit higher radio luminosities, with a median of $10^{24.6} ~\rm W Hz^{-1}$, compared to compact radio galaxies, which have a median of $10^{23.4} ~\rm W Hz^{-1}$. 
This observation aligns with the findings in Figure 21 of \cite{gordon23}, indicating higher 3 GHz luminosities for extended sources compared to single-component ones.
The CatWISE W1 magnitudes are similar, with median values of 15.2 and 15.4 for extended and compact radio galaxies, respectively. For a random sample of CatWISE sources, the median W1 magnitude is 16.8, underscoring that the infrared counterparts of radio galaxies are brighter than other infrared sources.
Additionally, data points for FR-I and FR-II type radio galaxies are plotted, revealing overlap between the two populations in luminosity, consistent with the findings in \cite{mingo19}. The median luminosities for FR-I and FR-II type radio galaxies are $10^{24.6} ~\rm W Hz^{-1}$ and $10^{24.3} ~\rm W Hz^{-1}$, respectively, while the median W1 magnitudes are both 15.2.

The middle panel of Figure~\ref{FIG:CounterpartStats2} shows contours on the AllWISE W1-W2 and W2-W3 color-color diagram for compact and extended radio galaxies. 
The AllWISE counterparts are selected as the closest CatWISE counterparts of radio galaxies with a maximum separation of $1^{\prime \prime}$. The plot displays contours at 5, 25, 50, 75, and 95 percentile levels for compact and extended radio galaxy counterparts.
Following the criteria outlined in \cite{mingo16} for identifying infrared host galaxies, we find a 63\% and 58\% AGN population (W1-W2>0.5) when a radio galaxy flux limit of 5 mJy is applied and AllWISE magnitudes above a signal-to-ratio larger than 3 are chosen. Applying the same flux and magnitude cuts, passive elliptical galaxies (W1-W2<0.5 and W2-W3<1.6) serve as hosts for 5\% and 16\% of compact and extended radio galaxies, respectively. Star-forming galaxies (W1-W2<0.5 and 1.6<W2-W3<3.4) host 10\% and 12\% of compact and extended radio galaxies, respectively. Ultra-luminous infrared galaxies are hosts to 14\% and 22\% of compact and extended radio galaxies. These proportions align with \cite{gordon23}, who find a majority of AGN hosts for compact and extended radio galaxies in the 3 GHz VLASS survey.
Additionally, data points for FR-I and FR-II type radio galaxies are shown, indicating substantial overlap between the two categories, consistent with the findings of \cite{mingo19}.

The bottom panel of Figure~\ref{FIG:CounterpartStats2} illustrates the color-color diagram (g-r) versus (r-i) for DES counterparts. The plot depicts distributions for compact and extended radio galaxy counterparts with contours at 5, 25, 50, 75, and 95 percentile levels.
The hosts exhibit consistent colours, with median values for (g-r, r-i) of (0.94, 0.74) and (1.2, 0.8) for compact and extended radio galaxies, respectively. For a random sample of DES sources, the median (g-r, r-i) colours are 3.7 and 1, indicating that optical hosts for radio galaxies have distinct colours compared to other optical sources.
As before, we also plot data points for FR-I and FR-II type radio galaxies, revealing significant overlap in their host galaxy colours.

\section{Summary}
\label{SEC:Summary}
With ongoing and upcoming experiments poised to identify tens of millions of radio galaxies, it becomes crucial to develop more efficient pipelines for constructing radio galaxy catalogues. 
This is essential to fully harness the potential insights offered by these observations and reduce the considerable human effort currently required.
In this study, we introduce a comprehensive detection pipeline designed to build radio galaxy catalogues through the application of cutting-edge computer vision algorithms. 
The pipeline employs a two-step process: initially detecting compact and extended radio galaxies along with their potential host galaxies in both radio and infrared images using computer vision networks, and subsequently utilizing the network predictions in the second stage to construct a catalogue. 
Leveraging this detection and catalogue construction pipeline, we create the first catalogue of radio galaxies for the pilot survey of the Evolutionary Map of the Universe (EMU-PS), conducted with the Australian Square Kilometer Array Pathfinder (ASKAP) telescope.

In developing the detection pipeline, we build upon the GalDINO network initially introduced in RadioGalaxyNET by \cite{gupta2023b} for radio galaxy detection. 
Expanding on the 2,800 FR-I, FR-II, FR-x, and R-type radio galaxies present in RadioGalaxyNET, we conduct visual inspections and label 2,090 compact radio galaxies and 99 sources with peculiar and other rare radio galaxies. 
This results in a dataset comprising approximately 5,000 radio galaxies, along with their corresponding infrared counterparts, utilized for training and evaluating the network.
The network is specifically trained to predict both the categories and bounding boxes for the radio galaxies, as well as the keypoint positions of their potential infrared host galaxies. 
Evaluation metrics are based on the Average Precision (AP) at a specified Intersection over Union (IoU) concerning the ground truth. We achieve an AP$_{50}$ of 73.2\% for the radio galaxy bounding box predictions and 71.7\% for the infrared host keypoint positions when considering the combined validation and test datasets. These results encompass predictions for all radio galaxy instances in each image. However, focusing solely on the radio galaxy located in the central part of the image reveals that 99\% of the central radio galaxy predictions exhibit an Intersection over Union (IoU) above 0.5 with respect to the ground truth bounding box in the evaluation set. 
For keypoint detection in the evaluation set, we observe that 98\% of keypoint locations have $<3^{\prime \prime}$ separation from the ground truth CatWISE infrared galaxy.
This observation suggests a strong alignment between the predicted bounding boxes and keypoints with the ground truth for the majority of the radio galaxies.

The pipeline for constructing the catalogue utilizes predictions from the Gal-DINO model for radio and infrared cutouts generated using the \textit{Selvay}-based components catalogue. 
This process results in 220,102 cutouts, for which bounding box, category, and keypoint predictions for the central source in the image are employed to assemble the consolidated catalogue.
Catalogue construction relies on the prediction confidence scores from the GalDINO network. 
For the 220,102 cutouts, the central source with the highest score is added to the consolidated catalogue first and is subsequently removed from the \textit{Selavy} catalogue. 
In the case of an extended radio galaxy, multiple components are grouped and added to the consolidated catalogue, while for a compact radio galaxy, one component is added.
The final catalogue comprises a total of 211,625 radio galaxies, including 201,211 compact radio galaxies and 10,414 extended radio galaxies. 
Among the extended radio galaxies, 582 are classified as FR-I, 5,602 as FR-II, 1,494 as FR-x, and 2,375 as R radio galaxies and 361 as Pec morphologies. Along with the GalDINO predictions, encompassing RA, DEC, peak flux density, integrated flux density, galaxy category, confidence score, etc., we also include corresponding \textit{Selavy} component identifiers in the catalogue.

Subsequently, a cross-matching process is employed to match radio galaxies with infrared (CatWISE) and optical catalogues (DES DR2, DESI Legacy Surveys DR8 and SuperCosmos) to identify their multi-wavelength counterparts. 
We observe that 75\% of the radio galaxies have counterparts in the CatWISE catalogue within a 3$^{\prime \prime}$ search radius. 
Further cross-matching of these CatWISE counterparts with optical catalogues, using a 2$^{\prime \prime}$ search radius, reveals that 64\%, 37\%, and 11\% of radio galaxies are matched with DES, DESI, and SCos catalogues, respectively.
The photometric redshifts obtained from the DESI legacy survey indicate a median redshift of 0.46 and 0.53 for compact and extended radio galaxies with a confidence score above 0.5. The farthest compact and extended radio galaxies are found at redshifts of 1.79 and 1.26, respectively.
Future research should concentrate on devising improved cross-matching methodologies to identify infrared and optical counterparts for the potential host positions detected by the computer vision pipeline. 
Subsequent research efforts should leverage the EMU-PS catalogue and the computer vision methodologies developed for cataloguing radio galaxies and their potential infrared hosts in ongoing and future radio surveys.
The existence of this catalogue will enable the advancement of more sophisticated machine-learning techniques tailored for the detection of radio galaxies in the next generation of radio surveys, including the EMU main survey.
To augment the variety of training data across the entirety of the southern sky, future studies should investigate the potential integration of active learning with a human-in-the-loop approach for the EMU main survey. 
This exploration should aim to enhance the diversity and quality of training data, contributing to the improvement of applied machine-learning methodologies.

\section*{Data Availability}
The catalogue is accessible at \url{https://data.csiro.au/collection/62055}\footnote{If access is restricted due to an embargo period, please contact the lead author for a copy.}, and the Gal-DINO network can be found at \url{https://github.com/Nikhel1/Gal-DINO}.

\section*{Acknowledgements}
We thank Ian Smail for his kind remarks on the draft and Matt Whitting for clarifying the purpose of the \textit{Selavy} source finder.
The Australian SKA Pathfinder is part of the Australia Telescope National Facility, which is managed by CSIRO. The operation of ASKAP is funded by the Australian Government with support from the National Collaborative Research Infrastructure Strategy. ASKAP uses the resources of the Pawsey Supercomputing Centre. The establishment of ASKAP, the Murchison Radio-astronomy Observatory and the Pawsey Supercomputing Centre are initiatives of the Australian Government, with support from the Government of Western Australia and the Science and Industry Endowment Fund. We acknowledge the Wajarri Yamatji people as the traditional owners of the Observatory site.
This publication makes use of data products from the Wide-field Infrared Survey Explorer, which is a joint project of the University of California, Los Angeles, and the Jet Propulsion Laboratory/California Institute of Technology, funded by the National Aeronautics and Space Administration.
The DESI Legacy Survey catalogues used in this paper were produced thanks to funding from the U.S. Department of Energy Office of Science and Office of High Energy Physics via grant DE-SC0007914.
The DES data management system is supported by the National Science Foundation under Grant Numbers AST-1138766 and AST-1536171.
NG acknowledges support from CSIRO’s Machine Learning and Artificial Intelligence Future Science Impossible Without You (MLAI FSP IWY) Platform.

\bibliography{ASKAP_PASA}

\appendix

\begin{table*}[!htb]
  \centering
  \begin{adjustbox}{width=24cm, angle=-90}
    \begin{tabular}{lrrrrrllrrlr}
\toprule
     EMU\_PS\_source\_name &  EMU\_PS\_ra\_deg &  EMU\_PS\_dec\_deg &  EMU\_PS\_flux\_peak &  EMU\_PS\_flux\_int &  EMU\_PS\_img\_id & EMU\_PS\_filename &                                  EMU\_PS\_radio\_bbox &  EMU\_PS\_ra\_cen\_bbox &  EMU\_PS\_dec\_cen\_bbox & EMU\_PS\_type &  EMU\_PS\_score \\
\midrule
EMU PS J203428.9-583922 &     308.620463 &      -58.656133 &             0.111 &            0.406 &        20636.0 &  J203428-583922 & [114.95496082305908, 112.12438583374023, 18.077... &          308.625802 &           -58.660541 &       FR-II &      0.336232 \\
EMU PS J210450.0-544653 &     316.208180 &      -54.781628 &             0.520 &            1.759 &        44380.0 &  J210449-544651 & [110.23069763183594, 116.73151588439941, 33.701... &          316.215076 &           -54.782720 &       FR-II &      0.875851 \\
EMU PS J202025.6-604829 &     305.106650 &      -60.808259 &             6.873 &            7.117 &        61989.0 &  J202025-604829 & [117.83510208129883, 117.61597347259521, 18.211... &          305.109103 &           -60.809591 &           C &      1.000000 \\
EMU PS J201051.4-604911 &     302.714042 &      -60.819875 &             1.659 &            2.803 &        51669.0 &  J201051-604912 & [117.64610290527344, 117.49090385437012, 18.175... &          302.717080 &           -60.821620 &           R &      0.926810 \\
EMU PS J221710.3-591718 &     334.293056 &      -59.288429 &             0.207 &            0.230 &       213454.0 &  J221710-591718 & [119.02629280090332, 118.01558828353882, 6.9277... &          334.294158 &           -59.289623 &           C &      1.000000 \\
EMU PS J211549.3-530854 &     318.955541 &      -53.148382 &             0.233 &            0.353 &       122337.0 &  J211549-530854 & [118.35806655883789, 117.77958583831787, 10.778... &          318.957050 &           -53.149657 &           C &      1.000000 \\
EMU PS J203404.9-622459 &     308.520235 &      -62.416623 &             0.188 &            0.175 &        22786.0 &  J203404-622500 & [118.08712196350098, 118.56851553916931, 6.6860... &          308.522591 &           -62.417530 &           C &      1.000000 \\
EMU PS J220545.6-502550 &     331.440087 &      -50.430620 &             0.246 &            0.213 &       204217.0 &  J220545-502550 & [117.54639053344727, 118.05950021743774, 10.523... &          331.442380 &           -50.431656 &           C &      1.000000 \\
EMU PS J220957.8-493722 &     332.490818 &      -49.622996 &             0.276 &            0.584 &       204538.0 &  J220957-493722 & [115.81414604187012, 117.93211603164673, 16.619... &          332.494258 &           -49.624156 &           C &      1.000000 \\
EMU PS J215754.2-561007 &     329.476005 &      -56.168620 &             0.167 &            0.328 &       195570.0 &  J215754-561007 & [118.29372668266296, 118.14089679718018, 7.9866... &          329.477719 &           -56.169769 &           C &      1.000000 \\
EMU PS J203010.8-513900 &     307.544833 &      -51.650116 &             0.217 &            0.191 &        70334.0 &  J203010-513900 & [117.80530834197998, 117.97474813461304, 8.0542... &          307.546795 &           -51.651312 &           C &      1.000000 \\
EMU PS J215126.8-541420 &     327.861737 &      -54.238876 &             0.268 &            0.223 &       174533.0 &  J215126-541420 & [118.01795673370361, 117.92996311187744, 8.1026... &          327.863568 &           -54.240122 &           C &      1.000000 \\
EMU PS J214906.1-583402 &     327.275246 &      -58.567375 &             0.209 &            0.190 &       182097.0 &  J214906-583402 & [118.46185970306396, 117.84657740592957, 7.7932... &          327.276874 &           -58.568656 &           C &      1.000000 \\
EMU PS J211916.8-544326 &     319.819851 &      -54.723969 &             0.183 &            0.157 &       130359.0 &  J211916-544326 & [117.97065353393555, 117.73301839828491, 8.4903... &          319.821789 &           -54.725227 &           C &      1.000000 \\
EMU PS J205252.6-621232 &     313.219070 &      -62.208944 &             1.161 &            1.222 &        94012.0 &  J205252-621232 & [117.86422109603882, 117.90887069702148, 12.471... &          313.221675 &           -62.210137 &           C &      1.000000 \\
EMU PS J210642.5-583558 &     316.676931 &      -58.599705 &             0.165 &            0.152 &        12070.0 &  J210644-583606 & [126.19098091125488, 124.15613746643066, 6.5276... &          316.677563 &           -58.599466 &           C &      1.000000 \\
EMU PS J221750.1-581022 &     334.458688 &      -58.172931 &             3.741 &            4.288 &       216888.0 &  J221750-581022 & [118.70282363891602, 117.93883609771729, 18.323... &          334.459988 &           -58.174079 &           C &      1.000000 \\
EMU PS J215806.8-562514 &     329.528227 &      -56.420644 &             0.231 &            0.199 &        21105.0 &  J215806-562514 & [117.8923499584198, 117.71990990638733, 7.67678... &          329.530333 &           -56.421999 &           C &      1.000000 \\
EMU PS J221455.8-552441 &     333.732565 &      -55.411415 &             0.413 &            0.430 &       210451.0 &  J221455-552441 & [118.31830191612244, 118.4164228439331, 7.40387... &          333.734221 &           -55.412391 &           C &      1.000000 \\
EMU PS J212155.8-575239 &     320.482449 &      -57.877519 &             0.288 &            0.281 &       141724.0 &  J212155-575239 & [117.8070297241211, 117.78631591796875, 9.70790... &          320.484654 &           -57.878780 &           C &      1.000000 \\
EMU PS J203015.8-591222 &     307.565981 &      -59.206349 &             0.234 &            0.258 &        71133.0 &  J203015-591223 & [118.66507339477539, 117.88196110725403, 10.023... &          307.567424 &           -59.207586 &           C &      1.000000 \\
EMU PS J211618.0-512748 &     319.075069 &      -51.463386 &             0.330 &            0.364 &       131778.0 &  J211618-512747 & [115.11660957336426, 118.06864356994629, 18.079... &          319.079284 &           -51.464402 &           C &      1.000000 \\
EMU PS J211259.1-600829 &     318.246210 &      -60.141563 &             0.243 &            0.217 &       123736.0 &  J211259-600829 & [118.15774059295654, 117.77396488189697, 8.5382... &          318.248207 &           -60.142844 &           C &      1.000000 \\
EMU PS J202826.1-604529 &     307.108867 &      -60.758282 &             0.205 &            0.289 &        62377.0 &  J202826-604530 & [116.72769498825073, 118.83250427246094, 12.089... &          307.112608 &           -60.759018 &           C &      1.000000 \\
EMU PS J221430.0-514836 &     333.625206 &      -51.810203 &             0.218 &            0.204 &       212263.0 &  J221430-514837 & [117.8178162574768, 118.24855709075928, 9.61719... &          333.627162 &           -51.811290 &           C &      1.000000 \\
EMU PS J210804.4-505451 &     317.018247 &      -50.914193 &            13.437 &           14.509 &       117896.0 &  J210804-505451 & [117.99701309204102, 118.04446601867676, 20.253... &          317.020049 &           -50.915451 &           C &      1.000000 \\
EMU PS J211409.8-515337 &     318.540776 &      -51.893874 &             0.179 &            0.164 &       132740.0 &  J211409-515338 & [117.68650841712952, 118.54129004478455, 7.7397... &          318.542935 &           -51.894744 &           C &      1.000000 \\
EMU PS J200756.5-530904 &     301.985462 &      -53.151168 &             8.076 &            8.143 &         9767.0 &  J200756-530904 & [118.41742849349976, 117.75225353240967, 14.678... &          301.986993 &           -53.152482 &           C &      1.000000 \\
EMU PS J212726.1-504105 &     321.858769 &      -50.684715 &             0.187 &            0.189 &       152584.0 &  J212726-504104 & [118.18173289299011, 117.69401121139526, 7.6252... &          321.860455 &           -50.685835 &           C &      1.000000 \\
EMU PS J212137.0-573451 &     320.404374 &      -57.580932 &             0.153 &            0.155 &       142384.0 &  J212137-573451 & [118.03919696807861, 117.52001810073853, 6.4959... &          320.406315 &           -57.582319 &           C &      1.000000 \\
EMU PS J204815.5-513212 &     312.064480 &      -51.536927 &             2.612 &            2.710 &        92882.0 &  J204815-513212 & [118.11067390441895, 117.90410423278809, 16.658... &          312.066284 &           -51.538101 &           C &      1.000000 \\
EMU PS J210744.8-592847 &     316.936852 &      -59.479934 &             0.248 &            0.325 &       116473.0 &  J210744-592847 & [118.10080814361572, 117.07810974121094, 9.4212... &          316.938872 &           -59.481599 &           C &      1.000000 \\
EMU PS J201956.0-514134 &     304.983519 &      -51.692972 &             0.162 &            0.228 &         5371.0 &  J201956-514135 & [117.88599300384521, 118.1196858882904, 7.00655... &          304.985467 &           -51.694169 &           C &      1.000000 \\
EMU PS J221338.0-485036 &     333.408215 &      -48.843387 &             0.327 &            0.294 &       208109.0 &  J221338-485036 & [118.82139110565186, 117.69000482559204, 10.722... &          333.409257 &           -48.844799 &           C &      1.000000 \\
EMU PS J203815.3-562656 &     309.563721 &      -56.449092 &             0.133 &            0.107 &        21307.0 &  J203815-562657 & [117.93629121780396, 118.3635094165802, 6.70931... &          309.565979 &           -56.450335 &           C &      1.000000 \\
EMU PS J204010.6-550441 &     310.044114 &      -55.078117 &             0.171 &            0.132 &         2265.0 &  J204010-550441 & [118.27358317375183, 118.30192565917969, 6.6099... &          310.045782 &           -55.079108 &           C &      1.000000 \\
EMU PS J215753.6-572734 &     329.473450 &      -57.459568 &             0.188 &            0.157 &       188677.0 &  J215753-572734 & [117.75442290306091, 118.10854291915894, 6.8046... &          329.475758 &           -57.460705 &           C &      1.000000 \\
EMU PS J214544.6-532821 &     326.435674 &      -53.472672 &             0.277 &            0.234 &       167981.0 &  J214544-532821 & [117.7500581741333, 118.28327894210815, 9.50040... &          326.437763 &           -53.473692 &           C &      1.000000 \\
EMU PS J213251.6-614021 &     323.214937 &      -61.672620 &             0.371 &            0.414 &       147109.0 &  J213251-614021 & [117.76328086853027, 117.79255771636963, 11.658... &          323.217479 &           -61.673850 &           C &      1.000000 \\
EMU PS J211247.9-512452 &     318.199647 &      -51.414506 &             1.210 &            1.201 &        41271.0 &  J211247-512452 & [118.4541368484497, 117.79197216033936, 13.1691... &          318.201103 &           -51.415671 &           C &      1.000000 \\
EMU PS J220711.3-531742 &     331.797066 &      -53.295085 &             0.537 &            0.571 &        43772.0 &  J220711-531742 & [118.22983360290527, 117.41423892974854, 12.602... &          331.798815 &           -53.296713 &           C &      1.000000 \\
EMU PS J211025.2-583054 &     317.604824 &      -58.515047 &             0.479 &            0.564 &       116187.0 &  J211025-583054 & [118.09213209152222, 117.34321403503418, 11.870... &          317.607034 &           -58.516747 &           C &      1.000000 \\
EMU PS J205246.0-492527 &     313.191794 &      -49.424331 &             0.190 &            0.190 &         3244.0 &  J205246-492527 & [118.0800838470459, 117.6152024269104, 9.001293... &          313.193433 &           -49.425608 &           C &      1.000000 \\
EMU PS J215141.8-630058 &     327.924269 &      -63.016353 &             1.248 &            1.251 &        35709.0 &  J215141-630100 & [118.23075866699219, 117.89554595947266, 11.326... &          327.926391 &           -63.017563 &           C &      1.000000 \\
EMU PS J202618.9-495342 &     306.578549 &      -49.895248 &             0.303 &            0.307 &        65982.0 &  J202618-495343 & [118.34949398040771, 117.99264478683472, 9.5052... &          306.579963 &           -49.896406 &           C &      1.000000 \\
EMU PS J220537.1-562103 &     331.404684 &      -56.350946 &             0.169 &            0.198 &       196222.0 &  J220537-562103 & [118.79777383804321, 118.27504134178162, 6.5261... &          331.405883 &           -56.352003 &           C &      1.000000 \\
EMU PS J201136.4-511906 &     302.901827 &      -51.318543 &             0.250 &            0.249 &        42964.0 &  J201136-511906 & [118.1784029006958, 117.36068391799927, 9.32610... &          302.903384 &           -51.320053 &           C &      1.000000 \\
EMU PS J212505.1-512441 &     321.271076 &      -51.411403 &             0.247 &            0.226 &       145450.0 &  J212505-512441 & [117.92004442214966, 117.99269580841064, 8.7904... &          321.272916 &           -51.412606 &           C &      1.000000 \\
EMU PS J201310.7-595507 &     303.294705 &      -59.918609 &             0.574 &            0.630 &        52248.0 &  J201310-595507 & [118.01626682281494, 117.97951889038086, 9.3812... &          303.296935 &           -59.919799 &           C &      1.000000 \\
EMU PS J204920.5-505059 &     312.335252 &      -50.849959 &             0.252 &            0.254 &        39567.0 &  J204920-505100 & [118.33566617965698, 117.80205202102661, 8.3895... &          312.336741 &           -50.851282 &           C &      1.000000 \\
EMU PS J220721.9-520613 &     331.841331 &      -52.103756 &             0.160 &            0.133 &       199684.0 &  J220721-520613 & [117.89800643920898, 118.20407342910767, 7.6273... &          331.843171 &           -52.104831 &           C &      1.000000 \\
EMU PS J215029.7-544432 &     327.623697 &      -54.742377 &             0.256 &            0.260 &       175175.0 &  J215029-544433 & [117.90699768066406, 118.04502010345459, 8.7571... &          327.625727 &           -54.743608 &           C &      1.000000 \\
EMU PS J202631.0-511808 &     306.629325 &      -51.302330 &             0.639 &            0.764 &        70396.0 &  J202631-511808 & [117.53812599182129, 117.94629383087158, 13.716... &          306.631512 &           -51.303570 &           C &      1.000000 \\
EMU PS J214245.0-612951 &     325.687476 &      -61.497549 &             0.305 &            0.268 &       165197.0 &  J214245-612951 & [117.88447761535645, 118.0237307548523, 9.01193... &          325.689855 &           -61.498711 &           C &      1.000000 \\
EMU PS J210052.2-531754 &     315.217477 &      -53.298329 &             9.277 &           10.543 &       109040.0 &  J210052-531753 & [120.71579837799072, 119.96245193481445, 29.086... &          315.216803 &           -53.298217 &           C &      1.000000 \\
EMU PS J205751.9-585021 &     314.466389 &      -58.839298 &             0.305 &            0.296 &        25499.0 &  J205751-585022 & [118.21373271942139, 118.11100578308105, 8.5130... &          314.468352 &           -58.840508 &           C &      0.997752 \\
EMU PS J212126.7-603944 &     320.361152 &      -60.662476 &             0.584 &            0.565 &       140359.0 &  J212126-603945 & [118.39134216308594, 117.9210376739502, 11.2748... &          320.363006 &           -60.663717 &           C &      1.000000 \\
EMU PS J202534.8-495354 &     306.395133 &      -49.898501 &             0.187 &            0.169 &        65537.0 &  J202534-495354 & [118.53793358802795, 118.03246450424194, 6.5903... &          306.396378 &           -49.899671 &           C &      1.000000 \\
EMU PS J210955.7-503036 &     317.481980 &      -50.510250 &             0.425 &            0.372 &       125640.0 &  J210955-503036 & [117.95140361785889, 118.1082592010498, 10.1012... &          317.483723 &           -50.511327 &           C &      1.000000 \\
EMU PS J215707.8-513635 &     329.282643 &      -51.609875 &             0.212 &            0.199 &       185881.0 &  J215707-513636 & [118.18254566192627, 118.08116340637207, 8.1607... &          329.284381 &           -51.611101 &           C &      1.000000 \\
\bottomrule
\end{tabular}
  
  \end{adjustbox}
  \caption{First few rows of the consolidated catalogue, excluding the remaining columns described in Table~\ref{TAB:column_description1} for brevity.}
  \label{TAB:Catalog}
\end{table*}

\label{lastpage}
\end{document}